\documentclass[12pt,preprint]{aastex}
 
\newcommand{\be}{  \begin{eqnarray} }
\newcommand{\ee}{  \end{eqnarray} }

\newcommand{\spd}[2]{\frac{\partial^2 #1}{\partial #2^2}}

\def\be{\begin{equation}}
\def\ee{\end{equation}}
\def\bea{\begin{eqnarray}}
\def\eea{\end{eqnarray}}

\def\spose#1{\hbox to 0pt{#1\hss}}
\def\lta{\mathrel{\spose{\lower 3pt\hbox{$\mathchar"218$}}
     \raise 2.0pt\hbox{$\mathchar"13C$}}}
\def\gta{\mathrel{\spose{\lower 3pt\hbox{$\mathchar"218$}}
     \raise 2.0pt\hbox{$\mathchar"13E$}}}
\font\syvec=cmbsy10                        
\font\gkvec=cmmib10                        
\def\bnabla{\hbox{{\syvec\char114}}}       
\def\bphi{\hbox{{\gkvec\char30}}}          

\def\baromr{\bar{\omega}_r}
\def\baromz{\bar{\omega}_z}

\begin{document}

%
       \title{Epicyclic Oscillations of Fluid Bodies:\\ 
	      Newtonian Non-Slender Torus}
%

\author{Omer M. Blaes\altaffilmark{1}}
\email{blaes@physics.ucsb.edu}

\author{Eva \v{S}r\'amkov\'a\altaffilmark{2}}
\email{sram\_eva@centrum.cz}

\author{Marek A. Abramowicz\altaffilmark{3,4,2}}
\email{marek@fy.chalmers.se}

\author{W{\l}odek Klu\'zniak\altaffilmark{4,5}}
\email{wlodek@camk.edu.pl}

\and

\author{Ulf Torkelsson\altaffilmark{3}}
\email{torkel@fy.chalmers.se}

\altaffiltext{1}{Department of Physics, University of California, 
Santa Barbara, CA 93106}

\altaffiltext{2}{Institute of Physics, Silesian University in Opava,
Bezru\v{c}ovo n\'am. 13, 746 01 Opava, Czech Republic}

\altaffiltext{3}{Department of Physics, G\"oteborg University, 
S-412 96 G\"oteborg, Sweden}

\altaffiltext{4}{Copernicus Astronomical Centre, Bartycka 18,
PL-00-716 Warszawa, Poland}

\altaffiltext{5}{Institute of Astronomy, Zielona G\'ora University,
Wie\.za Braniborska, Lubuska 2, PL-65-265 Zielona G\'ora, Poland}

%
              \begin{abstract}
We study epicyclic oscillations of fluid tori around black holes (in
the Paczy{\'n}ski-Wiita potential), and derive exact analytic expressions
for their radial and vertical eigenfrequencies $\nu_r$ and $\nu_z$,
to second order accuracy in the width of the torus. We prove
that pressure effects make the eigenfrequencies {\it smaller} than those 
for free particles. However, the particular ratio $\nu_z/\nu_r = 3/2$, 
that is important for the theory of high frequency QPOs, occurs when 
the fluid tori epicyclic frequencies $\nu_r$, $\nu_z$ are about 15\% 
{\it higher} than the ones corresponding to free particles. Our results 
therefore suggest that previous estimates of black hole spins from QPOs 
have produced values that are too high.    
\end{abstract}

%
%
%
              \section{Introduction}
%

The Fourier power density spectra of X-ray variability in Galactic
black hole X-ray binaries often reveal pairs of high
frequency QPOs (e.g., \citealt{str01,rem06}).  \citet{klu01a,klu01} suggested
that these high frequency QPOs
are caused by a non-linear resonance between two global modes of
oscillations in an accretion flow in strong gravity (here we denote these
modes by $\delta r$, $\delta z$), and pointed out that the observed
frequencies are in a commensurate (3:2) ratio. This suggestion was developed
by them and collaborators into the ``QPO resonance model''. The model
uses the theory of small non-linear oscillations (e.g., \citealt{nay79}),
and attempts to explain many observational properties of QPOs in X-ray
binaries by deriving
them directly from the differential equations that describe two weakly
coupled, non-linear oscillators (for more information, see the
collection of reviews in \citealt{abr05}),
\bea
&\delta {\ddot r} + (\omega_r)^2 \delta r&= 
{\cal X}_r (\delta r, \delta{\dot r}, \delta z, \delta{\dot z}), 
\nonumber\\
&\delta {\ddot z} + (\omega_z)^2 \delta z&= 
{\cal X}_z (\delta r, \delta{\dot r}, \delta z, \delta{\dot z}). 
\label{toy}
\eea
The resonance model does not address, however, the actual accretion flow
structure or the specific modes of oscillation, information upon which
the detailed form of equations (\ref{toy}) depend.

One possibility is that radial pressure gradients
set up fluid tori in the accretion flow which can support discrete,
trapped hydrodynamic modes.  That oscillations of such tori might be an
interesting model for QPOs was first recognized by Rezzolla and his collaborators
(\citealt{zan03}; \citealt{rez03}; see also \citealt{lee04}, \citealt{rub05},
and \citealt{bla06}).  It is not
yet clear whether such tori provide a realistic model for the accretion flow
in the steep power law state \citep{rem06}, where high frequency QPOs are
observed.  Nor is it clear whether their hydrodynamic modes of oscillation
can exist in the presence of magnetorotational (MRI) turbulence.
Nevertheless, pressure supported ``inner tori'' do
appear to be an ubiquitous flow feature of nonradiative global simulations
of MRI turbulence in accretion flows \citep{haw02,dev03}.  An example
of such an inner torus is shown in Figure \ref{fig-mami-torus} \citep{mac06}.

If torus-like structures do exist in the steep power law state, global
epicyclic oscillations of these tori are almost certainly the most robust
modes, as their existence is derived from the properties of the external
spacetime, not the internal properties of the torus \citep{abr06}.
While the existence of these modes is independent of the properties
of the torus, their actual frequencies and eigenfunctions are not.
It is this issue which we wish to address in the present paper:
how the frequencies of epicyclic modes of thick
(nonslender) fluid tori differ from the epicyclic frequencies of test
particles. As we shall discuss later in this paper, this question is
of direct relevance for an accurate estimate of the black hole spin from
the measured QPO frequencies. In order to answer this question, we
calculate analytically eigenfrequencies and eigenfunctions of the
epicyclic modes of nonslender tori up to the second order in the
torus thickness.

At first sight, vertical and radial epicyclic modes would seem to be a
terrible choice for a resonance, as test particle orbits in Kerr spacetime
are fully separable and there is therefore no nonlinear coupling between
these motions.  However, tori behave like test particles only when they
are very slender.  \cite{klu02} recognized that for nonslender tori the
frequencies of the epicyclic modes would be modified by pressure. They
derived an approximate formula for the epicyclic frequencies, radial
$\omega_r$ and vertical $\omega_z$, of fluid tori,
\begin{equation}
\label{pressure-klu02}
(\omega_r)^2 = (\omega_r^0)^2 - A_r c_{s0}^2\,\,\,\,\,
{\rm and}\,\,\,\,\,
(\omega_z)^2 = (\omega_z^0)^2 - A_z c_{s0}^2.
\end{equation}
Here $\omega_r^0$ and $\omega_z^0$ are the radial and vertical
epicyclic frequencies for particles, $c_{s0}$ is the sound speed at the
torus center, and the coefficients $A_r$ and $A_z$ are (not exactly
specified) functions of the equation of state and the background
gravitational potential.  Numerical work by \cite{rub05} revealed
that $A_r > 0$ and $A_z > 0$.
In this paper we analytically calculate explicit forms of $A_r$ and
$A_z$.  Another motivation for the present work is that it is a necessary step
toward deriving an explicit form of equations (1).  In the slender torus
limit, there is no nonlinear coupling of the epicyclic modes, but the
pressure corrections of equations (2) may give rise to nontrivial couplings.

For simplicity we model general relativistic effects throughout this paper with
the pseudo-Newtonian potential of \citet{pac80}.  The mathematics of
nonslender tori is complicated, and a Newtonian calculation is a useful
first step before attempting the calculation in full Kerr geometry.
We will publish an extention of our results to the Kerr geometry separately,
in O. Straub et al. (2007, in preparation).  In any case, the exact analytic
results here will be useful for oscillatory mode identification in numerically
simulated Paczy{\'n}ski-Wiita flows, as already attempted by \citet{bur06}
and M. Bursa \& M. Machida (2007, in preparation).

This paper is organized as follows. In Section \ref{sec:newton}
we briefly review the equilibrium structure of tori, citing results
that we will need later. In Section \ref{sec:baroclinic} we
demonstrate that radial and vertical epicyclic modes exist for a
completely general, baroclinic slender torus.  In Section
\ref{sec:thick} we then restrict consideration to polytropic, constant
specific angular momentum tori and derive the lowest order pressure
corrections to the epicyclic mode frequencies (second order) and
eigenfunctions (first order). We discuss our results and present our
conclusions in Section \ref{sec:conclusions}.

%
              \section{Newtonian Slender Torus}
              \label{sec:newton}
%

Consider an axially symmetric, inviscid rotating fluid body with
toroidal topology in equilibrium in an external axially symmetric
gravitational field. The flow is stationary and its velocity only has
an azimuthal component, ${\bf v}=\Omega r \hat{\bphi}$. In the paper we 
use cylindrical coordinates $\{r,\phi,z\}$ for all
calculations. The gravitational field is described by the potential
$\Phi(r,z)$, which we assume possesses reflection symmetry: $\Phi(r,z)
= \Phi(r,-z)$.

Dynamical equilibrium requires that
\begin{equation}
\Omega^2{\bf r} = \frac{{\bnabla} p}{\rho} + {\bnabla}\Phi.
\label{eq:Euler1}
\end{equation}
Here $p$ is the pressure and $\rho$ is the density.
Of particular interest is the circle where the pressure has zero 
gradient.  We shall call this the equilibrium circle, as it
corresponds to a balance between centrifugal and gravitational 
forces, as one can verify by substituting ${\bnabla} p=0$ into equation 
(\ref{eq:Euler1}). It follows from this equation that the circle lies
in the equatorial plane at a distance $r_0$ where the rotational
velocity $\Omega_0$ and the specific angular momentum $\ell_0$ of the
flow have their test particle (``Keplerian'') values $\Omega_K(r_0)$ and
$\ell_K(r_0)$. Let us introduce the effective potential of a test
particle with specific angular momentum $\ell_0$, ${\cal U}\equiv\Phi
+ \ell_0^2/(2r^2)$. The equilibrium circle lies at its minimum and the
Euler equation (\ref{eq:Euler1}) can be rewritten as
\begin{equation}
\frac{\ell^2 - \ell_0^2}{r^3}\hat{\bf r} = 
\frac{{\bnabla} p}{\rho} + {\bnabla}{\cal U}.
\label{eq:Euler2}
\end{equation}
The full equilibrium structure can easily be derived from this
equation in cases where the pressure can be expressed as a function of
density alone (a barotropic or pseudo-barotropic equilibrium, e.g.
\citeauthor{tas78}~\citeyear{tas78}). In this case, it is possible to 
find a potential $H$
such that  ${\bnabla} H = {\bnabla} p/\rho$. The left-hand side of equation
(\ref{eq:Euler2}) can then be expressed as a gradient.  Moreover, since
this gradient has  only a radial component, the corresponding
potential is a function of $r$ only and also the angular momentum
$\ell$ must be function of $r$ only -- the specific angular momentum
of the flow is constant on cylinders.

We now assume that the equilibrium pressure and density obey a polytropic
relation: $p\propto\rho^{1+1/n}$. Let us define two potentials
\begin{equation}
H = \int\frac{dp}{\rho} = (n+1)\frac{p}{\rho}
\quad{\rm and}\quad
\Psi = - \int_{r_0}^r\frac{\ell^2(r')-\ell_0^2}{{r'}^3}dr'.
\end{equation}
Substituting these into equation (\ref{eq:Euler2}), we obtain the
Bernoulli equation in the form
\begin{equation}
{\cal U}+\Psi+(n+1)\frac{p}{\rho} = {\rm const}.
\end{equation}
The constant can be evaluated by considering the equation at the
equilibrium point. Then we find
\begin{equation}
\frac{p}{\rho} = \frac{p_0}{\rho_0}\left[1 - \frac{1}{n c_{\rm s0}^2}
\left({\cal U}-{\cal U}_0+\Psi\right)\right] \equiv
\frac{p_0}{\rho_0} f(r,z),
\end{equation}
Here $c_{\rm s0}^2\equiv(n+1)p_0/(n\rho_0)$, so that $c_{\rm s0}$ is the
adiabatic sound speed $c_{\rm s}$ evaluated at the equilibrium point if the
barotropic equilibrium happens to also be isentropic. The pressure and density
profiles are given by $\rho = \rho_0 f^n$  and $p=p_0 f^{n+1}$.
Surfaces of constant pressure and density coincide with surfaces of
constant $f$.

It is useful to examine the behavior of the function $f$ in the
vicinity of the equilibrium point. For this purpose let us express the 
coordinates $r$ and $z$ as $r=(1+x) r_0$ and $z=y r_0$. The equilibrium 
point corresponds to $x=y=0$. For small $x$ and $y$ we have
\begin{equation}
f = 1 - \frac{r_0^2}{2 n c_{\rm s0}^2} 
\left[\left(\spd{{\cal U}}{r}\right)_0 x^2 +
\left(\spd{{\cal U}}{z}\right)_0 y^2 -
\frac{2\ell_0}{r_0^3}\left(\frac{d\ell}{dr}\right)_0 x^2 \right].
\end{equation}
The derivatives of the effective potential come from its expansion.
The first derivatives are missing because the equilibrium point
corresponds to a minimum of the effective potential. The mixed second
derivatives vanish due to the reflection symmetry. The term containing
the derivative of the specific angular momentum comes from the
expansion of the potential $\Psi$.  The second derivatives of the
effective potential with respect to $r$ and $z$ give us radial 
and vertical epicyclic frequencies $\omega_r$ and $\omega_z$ in the 
equilibrium point. Let us express them as fractions of the
orbital angular velocity $\Omega_0$ at the equilibrium circle,
$\omega_r=\bar{\omega}_r\Omega_0$, 
$\omega_z=\bar{\omega}_z\Omega_0$. Then we find
\begin{equation}
f = 1 - \frac{1}{\beta^2}\left\{\left[\bar{\omega}_r^2 - 
\frac{2r_0}{\ell_0}\left(\frac{d\ell}{dr}\right)_0 \right] x^2 +
\bar{\omega}_z^2 y^2 \right\},
\label{eq:f}
\end{equation}
where $\beta^2 \equiv (2 n c_{\rm s0}^2)/(r_0^2\Omega_0^2)$. The surfaces
of constant pressure and density have elliptical or hyperbolic
cross-sections depending on the sign of the square bracket. In fact,
it is possible to express the radial epicyclic frequency using the
gradient of the test particle (``Keplerian'') specific angular momentum as 
$\bar{\omega}_r^2 = (2r_0/\ell_0)(d\ell_{\rm K}/dr)_0$. 
Therefore, when the specific angular momentum of the fluid increases
more slowly than ``Keplerian'' (with increasing $r$), the equipressure 
surfaces have elliptical shape and the equilibrium point corresponds 
to the center of the torus with maximal pressure and density. When the
increase is faster the equilibrium point is a cusp which corresponds
to a saddle point in the pressure and density profiles.

Assume that the specific angular momentum distribution $\ell(r)$ is 
such that the equilibrium circle is at the center of the torus. The
surface of the torus is the equipressure surface where $f=0$.
Generally it can be far from the center -- in the region where our
approximation (small $x$ and $y$) is not valid. However, it is clear
from equation (\ref{eq:f}) that our approximation is valid in the
whole torus if $\beta$ is of the same order as $x$ and $y$, i.e., when
the flow is highly supersonic. In this limit the torus becomes
infinitesimally slender and its equipressure surfaces have elliptic
shape. In the particular case of a constant specific angular momentum 
distribution, these ellipses have semiaxises in the ratio of the
epicyclic frequencies. Moreover, when the external gravitational field
is Newtonian, $\Phi\propto 1/(r^2+z^2)^{1/2}$, these ellipses become circles.
It follows that then the torus has a circular cross-section with radius
$\beta r_0$.

%
              \section{Epicyclic Modes of Baroclinic Slender Tori}
              \label{sec:baroclinic}
%

Keeping the assumption that the torus is made of an ideal fluid, we
allow it to have arbitrary specific angular momentum $\ell(r,z)$
and entropy distributions.  Surfaces of constant pressure and density
need not coincide, and rotation need not be constant on cylinders.
The equilibrium configuration must still satisfy equation
(\ref{eq:Euler2}), but now the right hand side of that equation cannot
necessarily be expressed as a gradient, and we do not bother to
attempt to solve this equation for a detailed equilibrium structure.
However, we still assume that the torus equilibrium is slender.

For simplicity, we restrict consideration to axisymmetric
perturbations of this equilibrium. The equations describing Eulerian
perturbations are those of mass conservation, momentum conservation,
and adiabatic flow:
\begin{equation}
{\partial\delta\rho\over\partial t}+{1\over r}{\partial\over\partial
r}(r\rho \delta v_r)+{\partial\over\partial z}(\rho\delta v_z)=0,
\end{equation}
\begin{equation}
{\partial\delta v_r\over\partial t}-2\Omega\delta v_{\phi} =
{-1\over\rho}{\partial\delta p\over\partial r}+{\delta\rho\over\rho^2}
{\partial p\over\partial r},
\label{rmom}
\end{equation}
\begin{equation}
{\partial\delta v_\phi\over\partial t}+{1\over r}\delta{\bf
v}\cdot{\bnabla} \ell=0,
\label{angmom}
\end{equation}
\begin{equation}
{\partial\delta v_z\over\partial t}={-1\over\rho}
{\partial\delta p\over\partial z}+{\delta\rho\over\rho^2}
{\partial p\over\partial z},
\label{zmom}
\end{equation}
and
\begin{equation}
{\partial\delta p\over\partial t}-c_{\rm
s}^2{\partial\delta\rho\over\partial t}
+\delta{\bf v}\cdot({\bnabla}p-c_{\rm s}^2{\bnabla}\rho)=0.
\label{adiabatic}
\end{equation}

Now, for a slender torus, the derivatives of $r$ in the continuity
equation become negligible, and we may write instead
\begin{equation}
{\partial\delta\rho\over\partial t}+{\partial\over\partial r}
(\rho \delta v_r)+{\partial\over\partial z}(\rho\delta v_z)=0.
\label{cont}
\end{equation}
Hence from now on we may consider all vectors as two dimensional in
$r$ and $z$, which may be treated as Cartesian coordinates as far as
all vector operations are concerned.  We consider possible modes in
which $\delta v_r$ and $\delta v_z$ are spatially constant.  Then
after differentiating equations (\ref{rmom}) and (\ref{zmom})
with respect to time, and using equations (\ref{angmom}),
(\ref{adiabatic}) and (\ref{cont}) to eliminate $\delta v_\phi$,
$\delta p$, and $\delta\rho$, respectively, we obtain
\begin{equation}
{\partial^2\delta{\bf v}\over\partial t^2}+
\hat{\bf r}{2\Omega\over r}\delta{\bf v}\cdot
{\bnabla}\ell=\delta{\bf v}\cdot{\bnabla}
\left({1\over\rho}{\bnabla}p\right).
\end{equation}
For a slender torus, we may write
\begin{equation}
\hat{\bf r}{2\Omega\over r}\delta{\bf v}\cdot{\bnabla}\ell=
{\hat{\bf r}\over r^3}\delta{\bf v}\cdot{\bnabla}(\ell^2-\ell_0^2)\simeq
{\hat{\bf r}\over r_0^3}\delta{\bf v}\cdot{\bnabla}(\ell^2-\ell_0^2)=
\hat{\bf r}\delta{\bf v}\cdot{\bnabla}\left({\ell^2-\ell_0^2\over
r_0^3}\right)\simeq
\delta{\bf v}\cdot{\bnabla}\left[{\hat{\bf r}\over
r^3}(\ell^2-\ell_0^2)\right].
\end{equation}
Hence from the equilibrium condition (\ref{eq:Euler2}), we finally
obtain
\begin{equation}
{\partial^2\delta{\bf v}\over\partial t^2}=-\delta{\bf v}\cdot{\bnabla}
({\bnabla}{\cal U})|_0.
\label{eqnewtharmonic}
\end{equation}
The radial and vertical components of this equation give us our modes:
\begin{equation}
{\partial^2\delta v_r\over\partial t^2}=-\left({\partial^2{\cal
U}\over\partial r^2}\right)_0\delta v_r=-\omega_r^2\delta v_r
\end{equation}
and
\begin{equation}
{\partial^2\delta v_z\over\partial t^2}=-\left({\partial^2{\cal
U}\over\partial z^2}\right)_0\delta v_z=-\omega_z^2\delta v_z.
\end{equation}

%

              \section{Behavior of Epicylic Modes for Thick Tori}
	      \label{sec:thick}
%

We now turn to the behavior of the epicyclic modes for thicker tori.
To keep things simple, we restrict consideration to polytropic
tori with constant specific angular momentum.  This is unlikely
to be the most physically relevant case, particularly as existing
global simulations of accretion flows with magnetorotational turbulence
lead to near-Keplerian distributions of angular momentum
\citep{dev03,mac06,mat07}.
However, the eigenvalue problem is
self-adjoint for constant specific angular momentum, and the
calculations are therefore far more
straightforward.  While there will presumably be quantitative differences
for more general angular momentum distributions, there will still be
pressure corrections to the modes which are likely to be comparable to
those we calculate here.  We discuss this issue further in Section 5.

Because the equilibrium
tori are axisymmetric and stationary, we assume that all perturbations
have azimuthal and time dependence $\propto\exp[i(m\phi-\omega t)]$.
The periodicity of the solution in $\phi$ requires integer values of
$m$.

\citet{pap84} have shown that the perturbation equations are simplest
when expressed in terms of the variable
\begin{equation}
W \equiv \frac{\delta p}{\rho(m\Omega - \omega)},
\label{eq:W}
\end{equation}
where $\delta p$ is the Eulerian perturbation in pressure and $\Omega$
is the local angular velocity of fluid in the equilibrium torus. The
Eulerian perturbation in fluid velocity is simply $\delta{\bf v} =
i{\bnabla} W$, so that $W$ is directly proportional to the perturbed
velocity potential. In terms of $W$, the linear perturbations of the
torus are governed by the equation
\be
{1\over r}{\partial\over\partial r}\left(rf^n{\partial W\over\partial 
r}\right)
+{\partial\over\partial z}\left(f^n{\partial W\over\partial z}\right)
-{m^2\over r^2}f^n W+{2n(\omega-m\Omega)^2\over\beta^2 r_0^2\Omega_0^2}
f^{n-1}W=0,
\ee
where
\be
\beta^2\equiv{2(n+1)p_0\over\rho_0r_0^2\Omega_0^2}
\ee
is our slender torus perturbation parameter, and subscript zero refers
to the pressure maximum of the torus.  The equipotential function $f$
is given by
\be
f=1-{2\over\beta^2r_0^2\Omega_0^2}({\cal{U}}-{\cal{U}}_0),
\ee
where
\be
{\cal{U}}=\Phi(r,z)+{\ell_0^2\over2r^2}.
\ee
We assume the external gravitational potential $\Phi(r,z)$ is
symmetric with respect to reflections about the equatorial plane, so
all odd $z-$derivatives of ${\cal{U}}$ vanish in the equatorial plane.

The Papaloizou-Pringle equation may be written in abstract operator
form as
\be
\hat{L}W=-2n(\bar{\omega}-m\bar{\Omega})^2W.
\label{eqppthick}
\ee 
Here $\bar{\omega}\equiv\omega/\Omega_0$ and $\bar{\Omega} \equiv
\Omega/\Omega_0$. Also,
\be
\hat{L}\equiv\beta^2r_0^2\left[f{\partial^2\over\partial r^2} +
\left({f\over r} + n{\partial f\over\partial
r}\right){\partial\over\partial r} + f{\partial^2\over
\partial z^2}+n{\partial f\over\partial z}{\partial\over\partial z}-
{m^2f\over r^2}\right]
\ee
is a linear operator which is self-adjoint in the inner product
\be 
<W_1|W_2>\equiv{1\over\beta^2 r_0^3}\int dr\int dz rf^{n-1}W_1^\star W_2,
\ee
where the double integral is taken over the torus cross-section
defined by $f\ge0$.

%
              \subsection{Slender Torus Limit}
%

Change variables from $r$ and $z$ to
\be 
\bar{x}\equiv{r-r_0\over\beta r_0}\,\,\,\,\,{\rm and}\,\,\,\,\,
\bar{y}\equiv{z\over\beta r_0}.
\ee 

Then, in the slender torus limit $\beta \rightarrow0$, the
equipotential function becomes
\be 
f^{(0)}=1-\bar{\omega}_r^2\bar{x}^2-\baromz^2\bar{y}^2,
\ee 
where we use a superscript $(0)$ to denote the slender torus limit.
Here $\bar{\omega}_r$ and $\baromz$ are again the radial and vertical
epicyclic frequencies at the pressure maximum, scaled with the angular
velocity $\Omega_0$:
\be
\bar{\omega}_r^2={1\over\Omega_0^2}\left({\partial^2{\cal{U}}\over
\partial r^2}\right)_0\,\,\,\,\,{\rm and}\,\,\,\,\,
\baromz^2={1\over\Omega_0^2}\left({\partial^2{\cal{U}}\over
\partial z^2}\right)_0.
\ee 

The slender torus limit of the Papaloizou-Pringle equation is
\be 
\hat{L}^{(0)}W^{(0)}+2n\bar{\sigma}_0^2W^{(0)}=0,
\label{eqPPslender}
\ee 
where $\bar{\sigma}_0\equiv\bar{\omega}^{(0)}-m$ and
\be 
\hat{L}^{(0)}=f^{(0)}{\partial^2\over\partial\bar{x}^2}+n{\partial 
f^{(0)} \over\partial\bar{x}} {\partial\over\partial\bar{x}} +
f^{(0)}{\partial^2\over\partial\bar{y}^2}+n{\partial f^{(0)}
\over\partial\bar{y}} {\partial\over\partial\bar{y}}
\ee
is the slender torus limit of the linear operator $\hat{L}$. This is
still a self-adjoint operator with respect to the inner product
\be 
<W_1|W_2>=\int d\bar{x}\int d\bar{y}(f^{(0)})^{n-1}W_1^\star W_2,
\label{ipslender}
\ee 
where the integrals are now taken over the domain where $f^{(0)}\ge0$.
Because of this fact, the eigenvalues $\bar{\sigma}_0^2$ are all real,
and the corresponding eigenfunctions $W^{(0)}$ form a complete
orthonormal set in which any regular function defined on the domain
$f^{(0)}\ge0$ may be expanded.

\citet{bla85} derived the full set of orthonormal modes explicitly for
point mass potentials in which $\bar{\omega}_r = \baromz = 1$.
\citet{bla06} have derived the lowest order eigenfunctions and
eigenfrequencies for the more general case considered here. The
simplest modes are given in Table 1.  We are interested in the
behavior of the two epicyclic modes ($J=1$ and 2 in Table 1) as the
torus becomes thicker.  In the slender torus limit,
\begin{equation}
W_1^{(0)}\equiv a_1 \bar{x} e^{i(m\phi - \omega_1^{(0)}t)}
\quad {\rm and} \quad
W_2^{(0)}\equiv a_2 \bar{y} e^{i(m\phi - \omega_2^{(0)}t)}.
\label{eq:GlobModes}
\end{equation}
These are modes in which the fluid velocity is constant on torus
cross-sections and entirely radial or vertical in the case of
$W_1^{(0)}$ or $W_2^{(0)}$, respectively. These modes correspond to
radial and vertical epicyclic oscillations of a test particle in a
circular orbit. The correspondence between the fluid modes and test
particle epicyclic oscillations can be confirmed by looking at the
case of a general potential. Substituting equation
(\ref{eq:GlobModes}) into the Papaloizou-Pringle equation
(\ref{eqPPslender}), we find that they are still solutions of the more
general problem.  The corresponding eigenfrequencies in the corotating
frame $(\omega-m\Omega_0)$ are indeed the radial and vertical
epicyclic frequencies $\omega_r$ and $\omega_z$ in the case of $W_r$
and $W_z$, respectively.

%
              \subsection{First Order Perturbation Theory}
%

We now expand the general eigenvalue problem for thick tori in a power
series in $\beta$:
\be 
W=W^{(0)}+\beta W^{(1)}+\beta^2 W^{(2)}+...
\ee 
\be 
\bar{\omega}=\bar{\omega}^{(0)}+\beta\bar{\omega}^{(1)}+
\beta^2\bar{\omega}^{(2)}+...
\ee 
\be 
f=f^{(0)}+\beta f^{(1)}+\beta^2 f^{(2)}+...
\ee 
\be 
\bar{\Omega}=1+\beta\bar{\Omega}^{(1)}+\beta^2\bar{\Omega}^{(2)}+...
\ee 
\be 
\hat{L}=\hat{L}^{(0)}+\beta \hat{L}^{(1)}+\beta^2 \hat{L}^{(2)}+...
\ee 

Then if we expand the Papaloizou-Pringle equation (\ref{eqppthick}) 
to first order in $\beta$, we obtain
\be 
\hat{L}^{(1)}W^{(0)}+\hat{L}^{(0)}W^{(1)}=-2n\bar{\sigma}_0^2W^{(1)}
-4n\bar{\sigma}_0(\bar{\omega}^{(1)}-m\bar{\Omega}^{(1)})W^{(0)}.
\label{eqppfirst}
\ee 
We now expand $W^{(1)}$ in our orthonormal basis of zeroth order
eigenfunctions,
\be 
W^{(1)}=\sum_j b_j W_j^{(0)},
\ee 
and take the inner product of equation (\ref{eqppfirst}) with a
particular eigenfunction $W_J^{(0)}$.  If $J$ labels the particular
mode of interest, then this gives us an equation for the first order
correction to the eigenfrequency:
\be 
\bar{\omega}^{(1)}=-{1\over4n\bar{\sigma}_0}\left[
<W^{(0)}|\hat{L}^{(1)}|W^{(0)}>-4n\bar{\sigma}_0m<W^{(0)}|\bar{\Omega}
^{(1)} W^{(0)}>\right].
\label{eqomega1}
\ee 
If $J$ labels a different mode from the one of interest, then we get
an equation for the complex coefficients in the first order
eigenfunction:
\be 
b_J={1\over2n(\bar{\sigma}_{0J}^2-\bar{\sigma}_0^2)}\left[
<W^{(0)}_J|\hat{L}^{(1)}|W^{(0)}>-4n\bar{\sigma}_0m
<W^{(0)}_J|\bar{\Omega}^{(1)}W^{(0)}>\right].
\label{eqbj}
\ee 

Now, for a constant specific angular momentum torus,
\be 
\bar{\Omega}={\ell_0\over r^2\Omega_0}={r_0^2\over r^2} = (1 +
\beta\bar{x})^{-2}.
\ee 
Hence, we find $\bar{\Omega}^{(1)}=-2\bar{x}$. The first order correction 
to the equipotential function is given by
\be 
f^{(1)}=-{\bar{x}^3\over3}\bar{\cal U}_{rrr}-\bar{x}\bar{y}^2\bar{\cal
U}_{rzz},
\ee 
where
\be 
\bar{\cal U}_{rrr}\equiv{r_0\over\Omega_0^2}\left({\partial^3{\cal
U}\over \partial r^3}\right)_0\,\,\,\,\, {\rm and} \,\,\,\,\,
\bar{\cal U}_{rzz}\equiv{r_0\over\Omega_0^2}\left({\partial^3{\cal
U}\over \partial r\partial z^2}\right)_0.
\ee 
Finally, the first order correction to $\hat{L}$ is given by
\be 
\hat{L}^{(1)}=f^{(1)}{\partial^2\over\partial\bar{x}^2}+\left(f^{(0)}+
n{\partial f^{(1)}\over\partial\bar{x}}\right)
{\partial\over\partial\bar{x}}
+ f^{(1)}{\partial^2\over\partial\bar{y}^2}+n{\partial
f^{(1)}\over\partial \bar{y}}{\partial\over\partial\bar{y}}.
\ee 

Hence for the radial epicyclic mode, $W^{(0)}_1=a_1\bar{x}$, we have
\be 
\hat{L}^{(1)}|W^{(0)}_1>=a_1\left[1-\left(\bar{\omega}_r^2+
n\bar{\cal U}_{rrr}\right)\bar{x}^2-\left(\baromz^2+n\bar{\cal
U}_{rzz}\right) \bar{y}^2\right]
\label{eql1w1}
\ee 
and
\be 
\bar{\Omega}^{(1)}|W^{(0)}_1>=-2a_1\bar{x}^2.
\label{eqom1w1}
\ee 
On the other hand, for the vertical epicyclic mode, $W^{(0)}_2 =
a_2\bar{y}$,
\be 
\hat{L}^{(1)}|W^{(0)}_2>=-2na_2\bar{\cal U}_{rzz}\bar{x}\bar{y}
\label{eql1w2}
\ee 
and
\be 
\bar{\Omega}^{(1)}|W^{(0)}_2>=-2a_2\bar{x}\bar{y}.
\label{eqom1w2}
\ee 
In both cases, we immediately see from equation (\ref{eqomega1}) that
the first order correction to the frequency vanishes.

For the radial epicyclic mode, equations (\ref{eqbj}), (\ref{eql1w1})
and (\ref{eqom1w1}) give three nonzero coefficients for the first
order corrections to the eigenfunction, corresponding to the $J=0$, 4,
and 5 eigenmodes of Table 1.  Adding these together, the resulting
radial epicyclic eigenfunction $W_1$ turns out to be given by
\begin{eqnarray} 
{W_1\over a_1}&=&\bar{x}+{\beta\over2\baromr^4[(2n+3)\baromz^2-
(n+1)\baromr^2]}
\Biggl\{(2\baromz^2-\baromr^2)\left(\bar{\cal U}_{rrr}
-\baromr^2\mp8m\baromr\right)+\baromr^2\bar{\cal U}_{rzz}\cr
& &+\left[(2\baromz^2-\baromr^2)\left(-\baromr^2\pm8mn\baromr
-n\bar{\cal U}_{rrr}\right)\pm8m\baromr\baromz^2
-\baromz^2\bar{\cal U}_{rrr}+\baromr^2\bar{\cal U}_{rzz}\right]
\baromr^2\bar{x}^2\cr
& &+\left[\baromz^2\bar{\cal U}_{rrr}
-\baromr^2\baromz^2\mp8m\baromr\baromz^2-(n+1)\baromr^2\bar{\cal
U}_{rzz}\right]
\baromr^2\bar{y}^2\Biggr\} + {\cal O}(\beta^2),
\label{eqwrad}
\end{eqnarray} 
where $a_1$ is the normalization constant from Table 1.

The vertical epicyclic mode is much simpler.  Equations (\ref{eqbj}),
(\ref{eql1w2}) and (\ref{eqom1w2}) give only one nonzero
coefficient, corresponding to the $J=3$ eigenmode in Table 1.  The
resulting eigenfunction is then given by
\begin{equation} 
{W_2\over a_2}=\bar{y}+{\beta\over\baromr^2}
\left(\pm4m\baromz-\bar{\cal U}_{rzz}\right)\bar{x}\bar{y}+{\cal
O}(\beta^2).
\label{eqwvert}
\end{equation} 

%
              \subsection{Second Order Perturbation Theory}
%

Because the first order frequency corrections vanish, we have to go 
to second order to determine the finite pressure corrections to the
epicyclic mode frequencies.

The second order expansion terms are $\bar{\Omega}^{(2)}=3\bar{x}^2$,
\be 
f^{(2)}=-{1\over12}\left(\bar{x}^4\bar{\cal U}_{rrrr}
+6\bar{x}^2\bar{y}^2\bar{\cal U}_{rrzz}+\bar{y}^4\bar{\cal U}_{zzzz}
\right),
\ee 
and
\be 
\hat{L}^{(2)}=f^{(2)}{\partial^2\over\partial\bar{x}^2}+\left(f^{(1)}-
\bar{x}
f^{(0)}+n{\partial f^{(2)}\over\partial\bar{x}}\right)
{\partial\over\partial\bar{x}}
+f^{(2)}{\partial^2\over\partial\bar{y}^2}+n{\partial
f^{(2)}\over\partial
\bar{y}}{\partial\over\partial\bar{y}}-m^2f^{(0)}.
\ee 
Here
\be 
\bar{\cal U}_{rrrr}\equiv{r_0^2\over\Omega_0^2}\left({\partial^4{\cal
U} \over\partial r^4}\right)_0,\,\,\,\,\,
\bar{\cal U}_{rrzz}\equiv{r_0^2\over\Omega_0^2}\left({\partial^4{\cal
U} \over\partial r^2\partial z^2}\right)_0,\,\,\,\,\,
{\rm and}\,\,\,\,\,
\bar{\cal U}_{zzzz}\equiv{r_0^2\over\Omega_0^2}\left({\partial^4{\cal
U} \over\partial z^4}\right)_0.
\ee 

The second-order terms in the Papaloizou-Pringle equation
(\ref{eqppthick}) give
\begin{eqnarray} 
\hat{L}^{(2)}W^{(0)}+\hat{L}^{(1)}W^{(1)}&+&\hat{L}^{(0)}W^{(2)}=
-2n\bar{\sigma}_0^2W^{(2)}-4n\bar{\sigma}_0
(\bar{\omega}^{(1)}-m\bar{\Omega}^{(1)})W^{(1)}\cr
& &-4n\bar{\sigma}_0(\bar{\omega}^{(2)}-m\bar{\Omega}^{(2)})W^{(0)}
-2n(\bar{\omega}^{(1)}-m\bar{\Omega}^{(1)})^2W^{(0)}.
\label{eqppsecond} 
\end{eqnarray} 

Once again, we expand the second order eigenfunction in terms of the
zeroth order eigenfunctions,
\be  
W^{(2)}=\sum_j c_jW^{(0)}_j.
\ee 
Then on taking the inner product of equation (\ref{eqppsecond}) with
the zeroth order eigenfunction of the mode of interest, we find that
the second order correction to the frequency is (for the case when the
first order correction $\bar{\omega}^{(1)}$ vanishes)
\begin{eqnarray} 
\bar{\omega}^{(2)}&=&-{1\over4n\bar{\sigma}_0}\Biggl\{
<W^{(0)}|[\hat{L}^{(2)}-4n\bar{\sigma}_0m\bar{\Omega}^{(2)}+2nm^2
(\bar{\Omega}^{(1)})^2]|W^{(0)}>\cr
& &+\sum_jb_j<W^{(0)}|[\hat{L}^{(1)}-4nm
\bar{\sigma}_0\bar{\Omega}^{(1)}]|W^{(0)}_j>\Biggr\}.
\label{eqomega2}
\end{eqnarray} 

For the radial epicyclic mode, we find
\begin{eqnarray} 
\bar{\omega}_1&=&\pm\baromr+m\mp\frac{\beta^2}{8(n+2)\baromr^5\baromz^
2} \Biggl\{-\baromr^2\baromz^2\bar{\cal U}_{rrrr}-\baromr^4\bar{\cal
U}_{rrzz} -2m^2\baromr^4\baromz^2+{1\over[(2n+3)\baromz^2-(n +
1)\baromr^2]}\times\cr &\times&\biggl\{
2\baromr^4\baromz^2[(n+2)\baromr^2-(2n+5)\baromz^2]+
2\baromr^2\baromz^2\bar{\cal U}_{rrr}[(n-1)\baromr^2-(2n-1)\baromz^2]
+4\baromr^4\baromz^2\bar{\cal U}_{rzz}\cr
& &+\baromz^2\bar{\cal U}_{rrr}^2[(5+6n)\baromz^2-(3n+1)\baromr^2]
+\baromr^2\bar{\cal U}_{rrr}\bar{\cal
U}_{rzz}[(2n-1)\baromz^2-(n+1)\baromr^2]
+2\baromr^4(n+1)\bar{\cal U}_{rzz}^2\cr
& &\pm4m\baromr^3\baromz^2[(7n+15)\baromr^2-(14n+37)\baromz^2]
\pm4m\baromr\baromz^2\bar{\cal
U}_{rrr}[(7n-1)\baromr^2-(14n+5)\baromz^2]\cr
& &\pm4m\baromr^3\bar{\cal U}_{rzz}[(n+1)\baromr^2-(2n-5)\baromz^2]
+8m^2\baromr^2\baromz^2[(9-7n)\baromr^2+(14n-11)\baromz^2]
\biggr\}\Biggr\}\cr
& &+{\cal O}(\beta^3).
\label{eqomrad}
\end{eqnarray} 

The frequency of the vertical epicyclic mode is
\begin{eqnarray} 
\bar{\omega}_2&=&\pm\baromz+m\mp{\beta^2\over8\baromr^4\baromz^3(n+2)}
\Biggl\{-\baromr^2\baromz^2\bar{\cal U}_{rrzz}-\baromr^4\bar{\cal
U}_{zzzz}
-2\baromr^2\baromz^2\bar{\cal U}_{rzz}\cr
& &+\bar{\cal U}_{rzz}[\baromz^2\bar{\cal U}_{rrr}
+(2\baromz^2+3\baromr^2)\bar{\cal U}_{rzz}]
\mp4m\baromz[\baromr^2\baromz^2+\baromz^2\bar{\cal U}_{rrr}
+(4\baromz^2+3\baromr^2)\bar{\cal U}_{rzz}]\cr
& &+2m^2\baromz^2(16\baromz^2+4\baromr^2-\baromr^4)\Biggr\}+
{\cal O}(\beta^3).
\label{eqomvert} 
\end{eqnarray} 

%
              \subsection{Spherically Symmetric Point Mass Potential}
%

In this case we have
\begin{eqnarray} 
\baromr^2&=&\baromz^2=1,\,\,\,\,\,\bar{\cal U}_{rrr}=-6,\,\,\,\,\,
\bar{\cal U}_{rzz}=-3,\cr
\bar{\cal U}_{rrrr}&=&36,
\,\,\,\,\,\bar{\cal U}_{rrzz}=12,\,\,\,\,\,{\rm and}\,\,\,\,\,
\bar{\cal U}_{zzzz}=-9.
\end{eqnarray} 
For the radial epicyclic mode,
\begin{equation} 
\bar{\omega}_1 = \pm 1+m \mp
{\beta^2\over4(n+2)^2}\left(53n+6\pm76nm\mp8m
+27nm^2-10m^2\right) + {\cal O}(\beta^3)
\label{omradnewt}
\end{equation} 
and
\begin{eqnarray} 
W_1=\left[{2n(n+1)\over\pi}\right]^{1/2}\Biggl\{\bar{x}&+&{\beta\over
n+2} \Biggl[-5\mp4m+(3n+1\pm4m\pm4nm)\bar{x}^2\cr
& &+\left({3\over2}n-2\mp4m\right)
\bar{y}^2\Biggr]\Biggr\}+{\cal O}(\beta^2).
\end{eqnarray} 
For the vertical epicyclic mode, 
\begin{equation}
\bar{\omega}_2=\pm1+m\mp{\beta^2\over4(n+2)}(33\pm52m+19m^2)+{\cal
O}(\beta^3)
\label{omvertnewt}
\end{equation} 
and
\begin{equation} 
W_2=\left[{2n(n+1)\over\pi}\right]^{1/2}\left[\bar{y}+\beta(3\pm4m)
\bar{x} \bar{y}\right]+{\cal O}(\beta^2).
\end{equation} 
It is perhaps interesting to note that for $n=3$ (i.e. radiation
pressure dominated) tori, the axisymmetric ($m=0$) radial and vertical
epicyclic modes frequencies continue to be degenerate to the $\beta^2$
order of accuracy.

%
              \subsection{Pseudo-Newtonian Potential}
%

In the case of the \citet{pac80} pseudo-Newtonian potential, in units
where $c=G=M=1$,
\begin{equation} 
\Phi=-{1\over(r^2+z^2)^{1/2}-2},
\end{equation} 
we have
\begin{eqnarray} 
\bar{\omega}_r^2&=&{r_0-6\over r_0-2},\,\,\,\,\,\baromz^2 = 1,
\,\,\,\,\,\bar{\cal U}_{rrr}=-{6(r_0^2-8r_0+8)\over(r_0-2)^2},\cr
\bar{\cal U}_{rzz}&=&-{3r_0-2\over r_0-2},\,\,\,\,\,\bar{\cal
U}_{rrrr}= {12(3r_0^3-30r_0^2+60r_0-40)\over(r_0-2)^3},\cr
\bar{\cal U}_{rrzz}&=&{4(3r_0^2-4r_0+2)\over(r_0-2)^2},\,\,\,\,\,{\rm
and} \,\,\,\,\,\bar{\cal U}_{zzzz}=-{3(3r_0-2)\over r_0-2}.
\end{eqnarray} 

These expressions may be substituted into equations (\ref{eqwrad}),
(\ref{eqwvert}), (\ref{eqomrad}) and (\ref{eqomvert}) to obtain rather
complicated, explicit expressions for the epicyclic mode
eigenfunctions and eigenfrequencies as a function of location of the
torus pressure maximum and $\beta$.  Rather than present those
expressions here, we use them to numerically evaluate the mode
frequencies. 

%
              \subsubsection{Axisymmetric Modes}
%

\begin{figure}
\plottwo{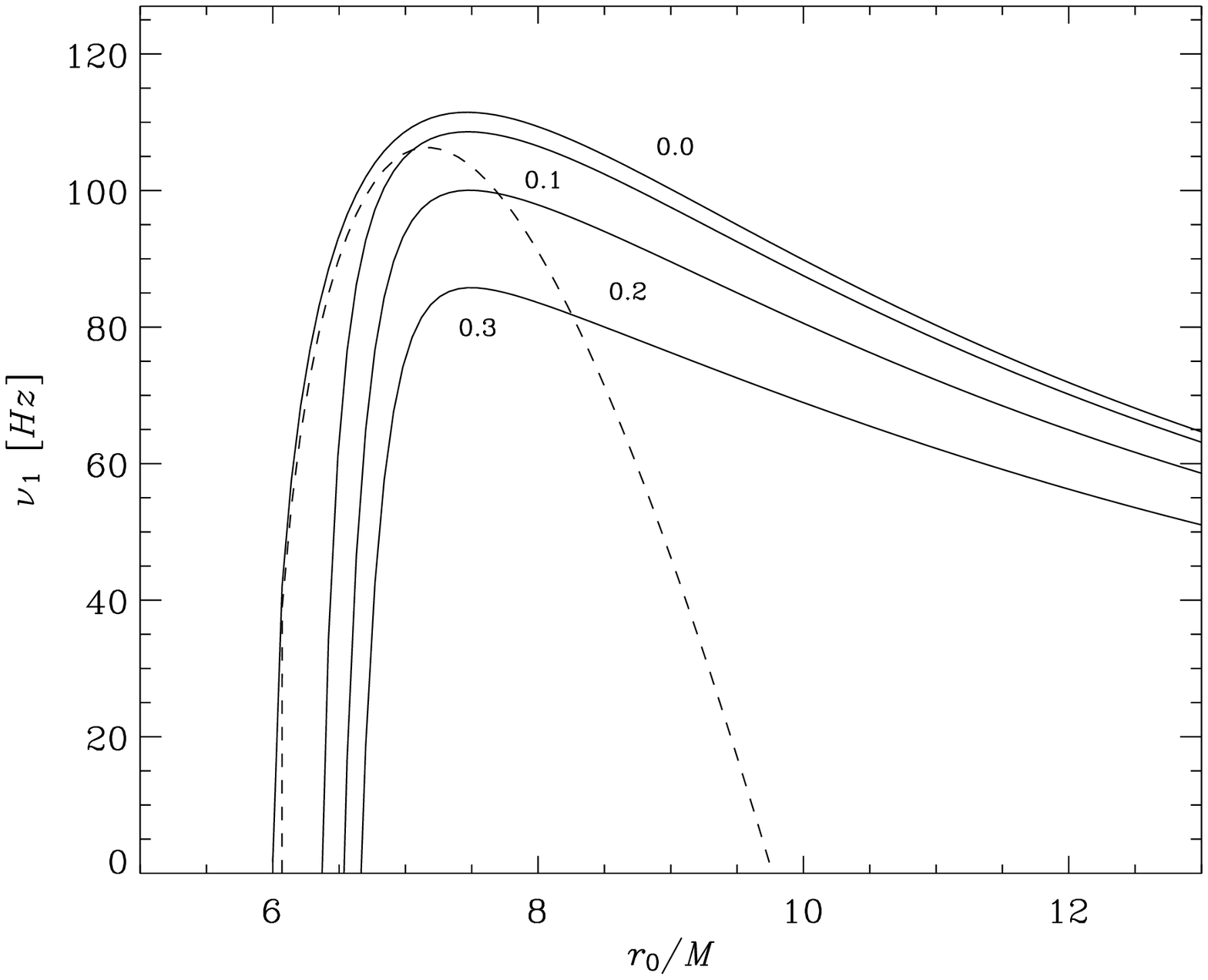}{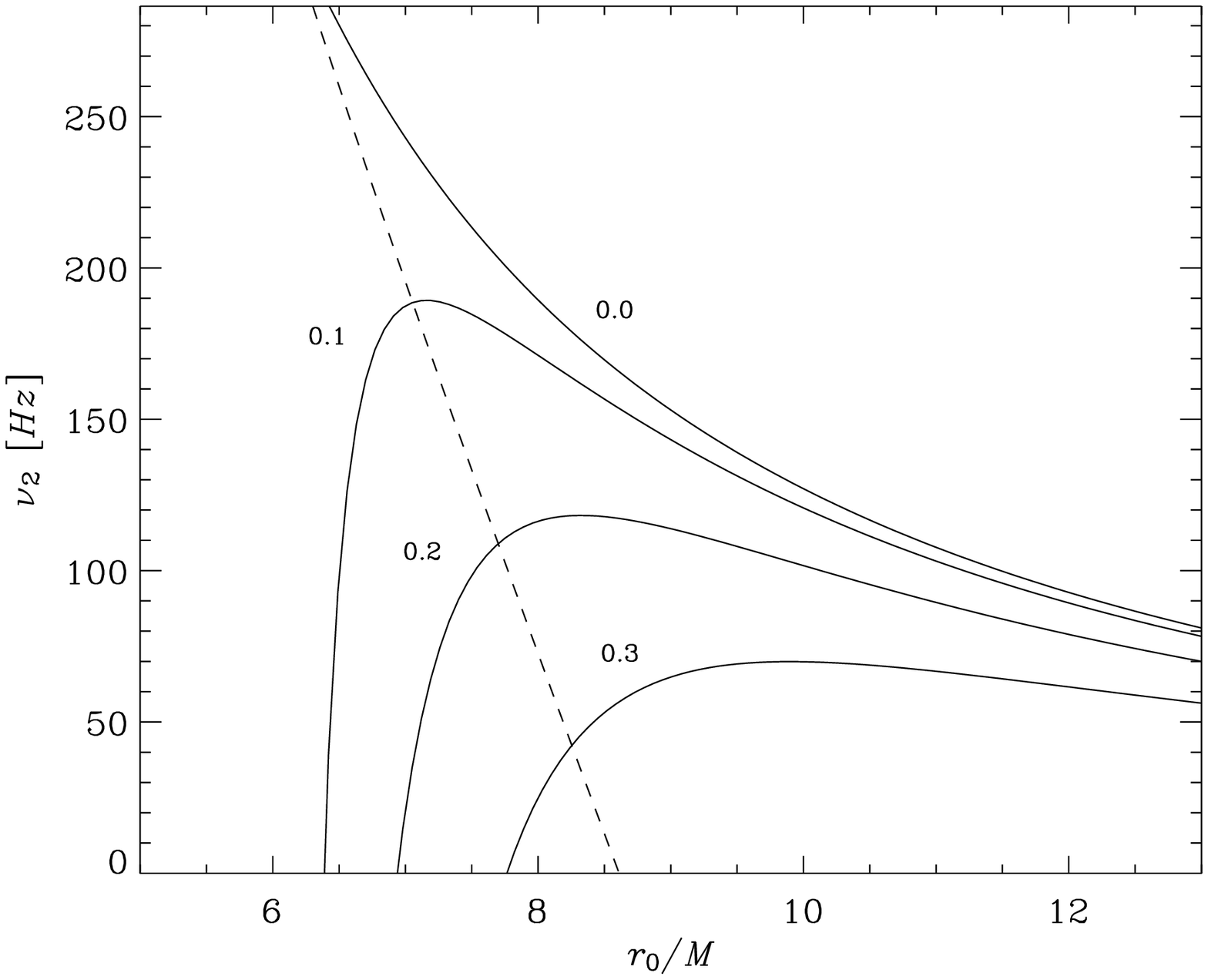}
\caption{ {\it Left:} Frequency of the axisymmetric ($m=0$) radial
epicyclic  mode, $\nu_1=\omega_1/(2\pi)$, for an $n=3$ torus in a
pseudo-Newtonian potential with black hole mass $M=10M_{\odot}$, as a
function of radius of the torus pressure maximum in units of the
gravitational radius.  Different solid curves correspond to
different values of the pressure parameter $\beta$, and are labeled as
such. Models below the dashed curve are unphysical, as the
corresponding tori extend outside the critical cusp equipotential.
{\it Right:} Same as the left Figure, but for the axisymmetric
vertical epicyclic mode. Once again, physical models lie above (to the
right) of the dashed line.}
\label{axisym}
\end{figure}

Figure \ref{axisym} shows the behavior of the axisymmetric ($m=0$)
radial and vertical epicyclic mode frequencies.  The $\beta=0$ curves
in these Figures correspond to the test particle frequencies.  For a
given pressure maximum radius $r_0$, there is a maximum value of $\beta$
beyond which equilibrium tori cannot exist. This limit is indicated by
the dashed curves in the Figures.

\begin{figure}
\plotone{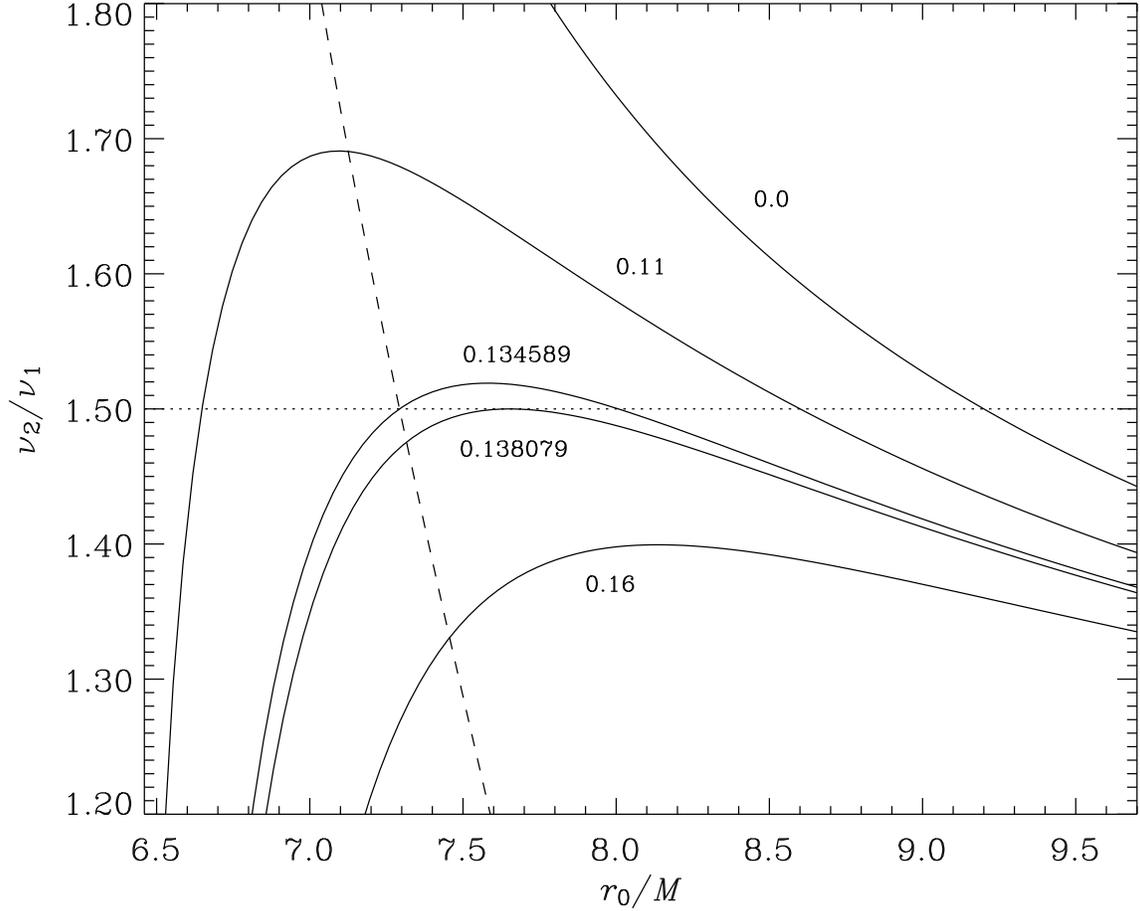}
\caption{Ratio $\nu_2/\nu_1=\omega_2/\omega_1$ of the vertical to
radial axisymmetric epicyclic mode frequencies for an $n=3$ torus in a
pseudo-Newtonian potential, for different values of $\beta$ and as a
function of radius of the torus pressure maximum. Again, physical
models lie above and to the right of the dashed line. The dotted line
indicates a 3:2 commensurability between the two modes. For a slender
torus with a particular value of $\beta<0.134589$, there is one radius
for which the axisymmetric mode frequency ratio will be 3:2. For
somewhat thicker tori with values of $\beta$ satisfying
$0.134589<\beta<0.138079\equiv\beta_{\rm max}$, there are two
locations of the torus pressure maximum where the mode frequencies are
in a 3:2 ratio.  No such radii exist for still thicker tori with
$\beta > \beta_{max}$.}
\label{ratio}
\end{figure}

\begin{figure}
\plottwo{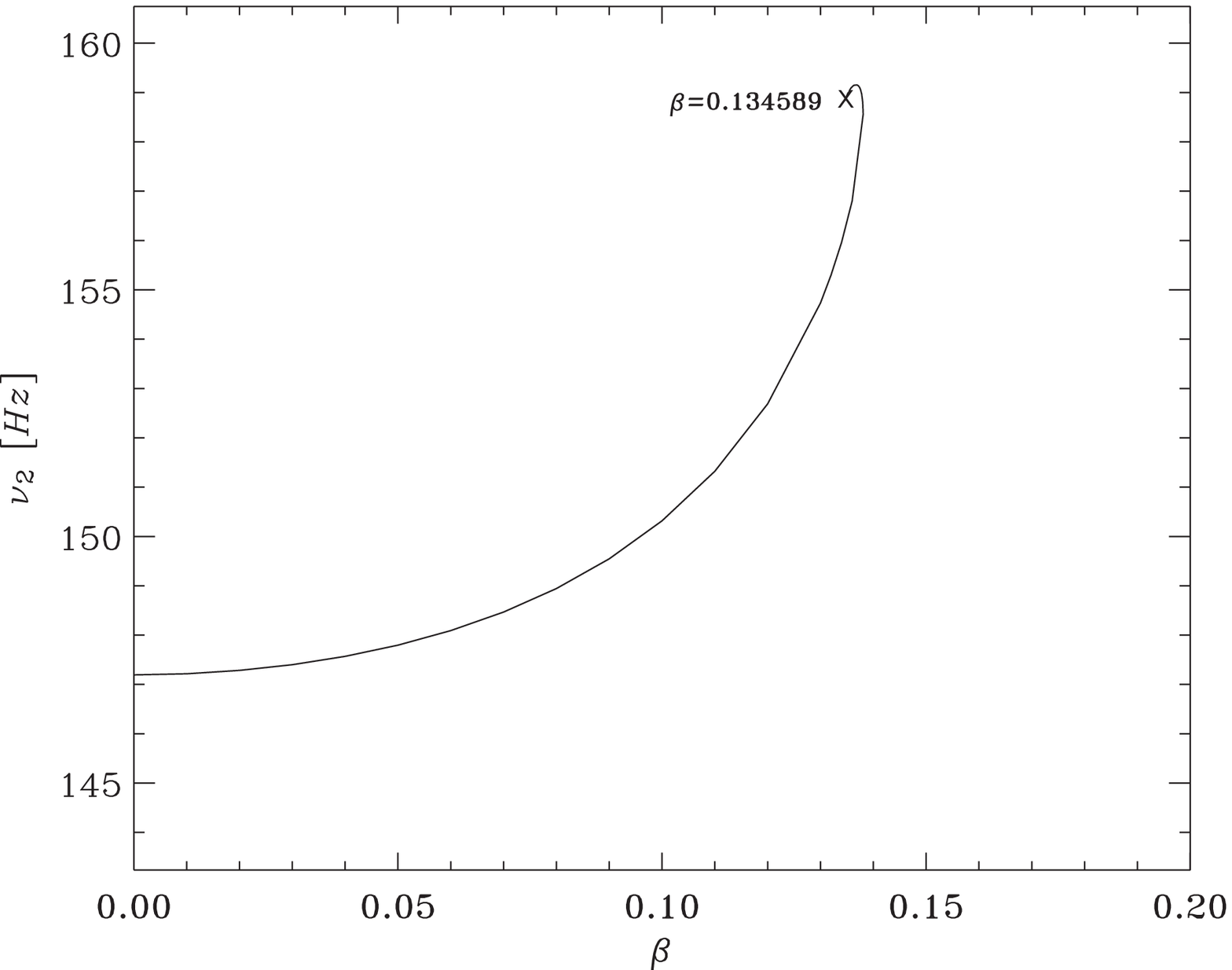}{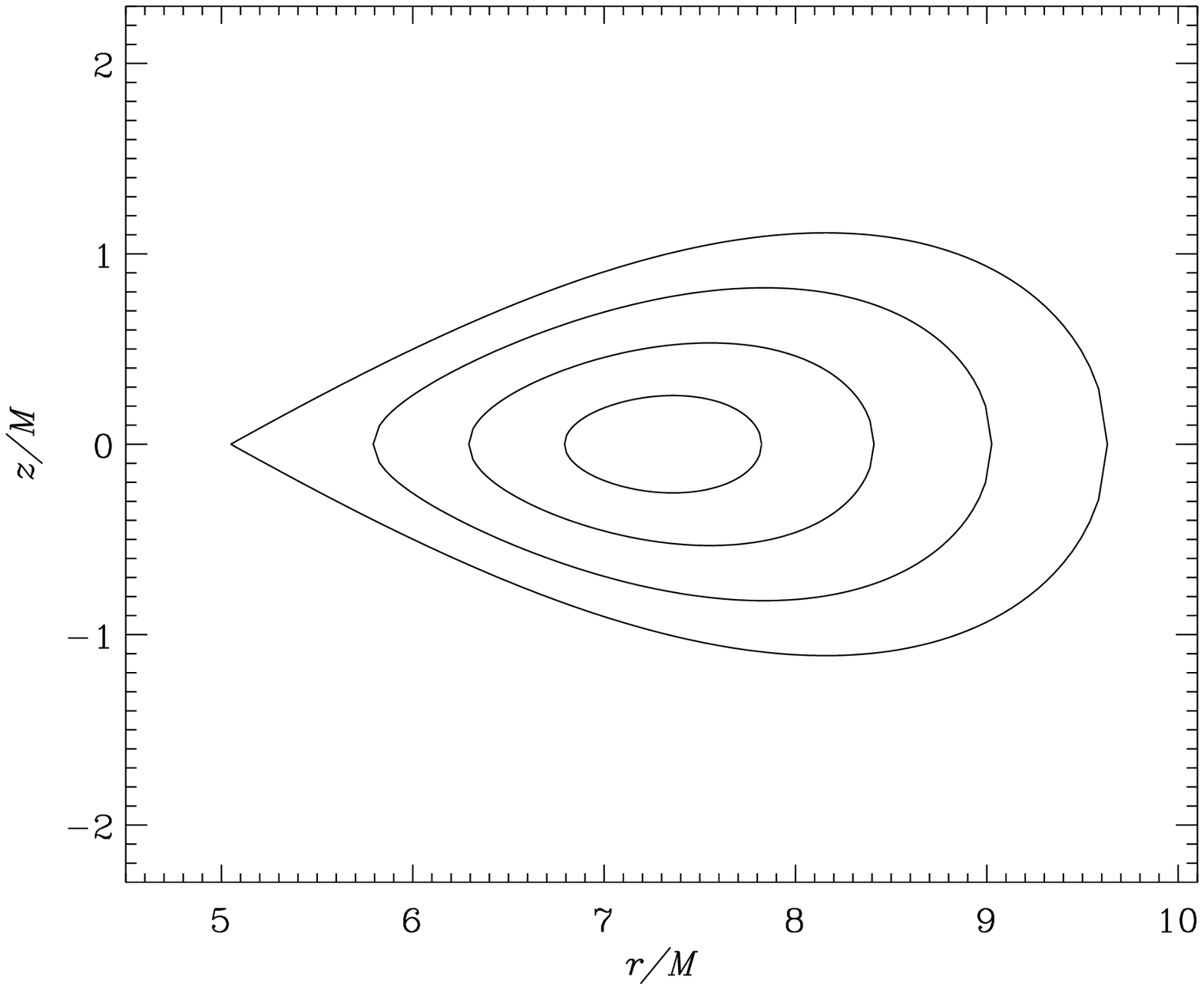}
\caption{{\it Left}:  Axisymmetric vertical epicyclic mode frequency
$\nu_2=\omega_2/(2\pi)$ for tori which have the axisymmetric vertical
and radial modes in a 3:2 ratio, plotted as a function of $\beta$.
Again, the tori are assumed to have $n=3$ and orbit in a
pseudo-Newtonian potential with $M=10M_{\odot}$.  The $\beta=0.134589$
point in the Figure corresponds to the intersection of the dashed and
dotted lines in Figure \ref{ratio}. {\it Right}:  Meridional
cross-section of a nonslender torus with  $\beta=0.134589$
and a cusp at $r=5.0485 M$. The torus is centered at $r=7.293 M$ where
the axisymmetric epicyclic mode frequencies $\omega_1$ and $\omega_2$
for this $\beta$ are in a 3:2 ratio. This case corresponds
to the crossing point of the dashed and dotted lines in Figure
\ref{ratio}.}
\label{3.2}
\end{figure}

Note from Figure \ref{axisym} that the axisymmetric vertical epicyclic
mode frequency for nonslender tori exhibits a maximum value as a
function of $r_0$, in contrast to the behavior of the test particle
frequency in a pseudo-Newtonian potential.  This gives rise to
interesting behavior of the ratio of the vertical to radial mode
frequencies, as shown in Figure \ref{ratio}. Once again, the dashed
line indicates the limit beyond which equilibrium tori can exist.
For small values of $\beta$ the ratio
of test particle frequencies rises monotonically toward smaller
radii, and there is therefore a unique radius at which the
frequency ratio can take on any specific value.  
The dotted line shows the 3:2 frequency ratio that is found in the high
frequency QPOs.  As the torus thickens, the radius at which the 3:2
commensurability occurs moves inward, and as shown in Figure \ref{3.2}
(the left part), each of the mode frequencies increases.  This
continues until $\beta=0.134589$, when  suddenly tori
at two different radii display 
a 3:2 commensurability between the axisymmetric epicyclic
modes.  The inner torus produces higher frequencies than the outer
torus. As $\beta$ increases further, these two tori move toward
each other, converging at $\beta=0.138079$. For still thicker tori,
there is no radius at which the axisymmetric epicyclic modes are in a
3:2 ratio. Hence we conclude that the axisymmetric epicyclic modes can
represent both observed high frequency QPOs only if the torus is not
too thick.  The right part of Figure \ref{3.2} shows the equipotential
surfaces of the torus that exhibits the highest such mode frequencies
while still retaining the 3:2 commensurability.

\begin{figure}
\plottwo{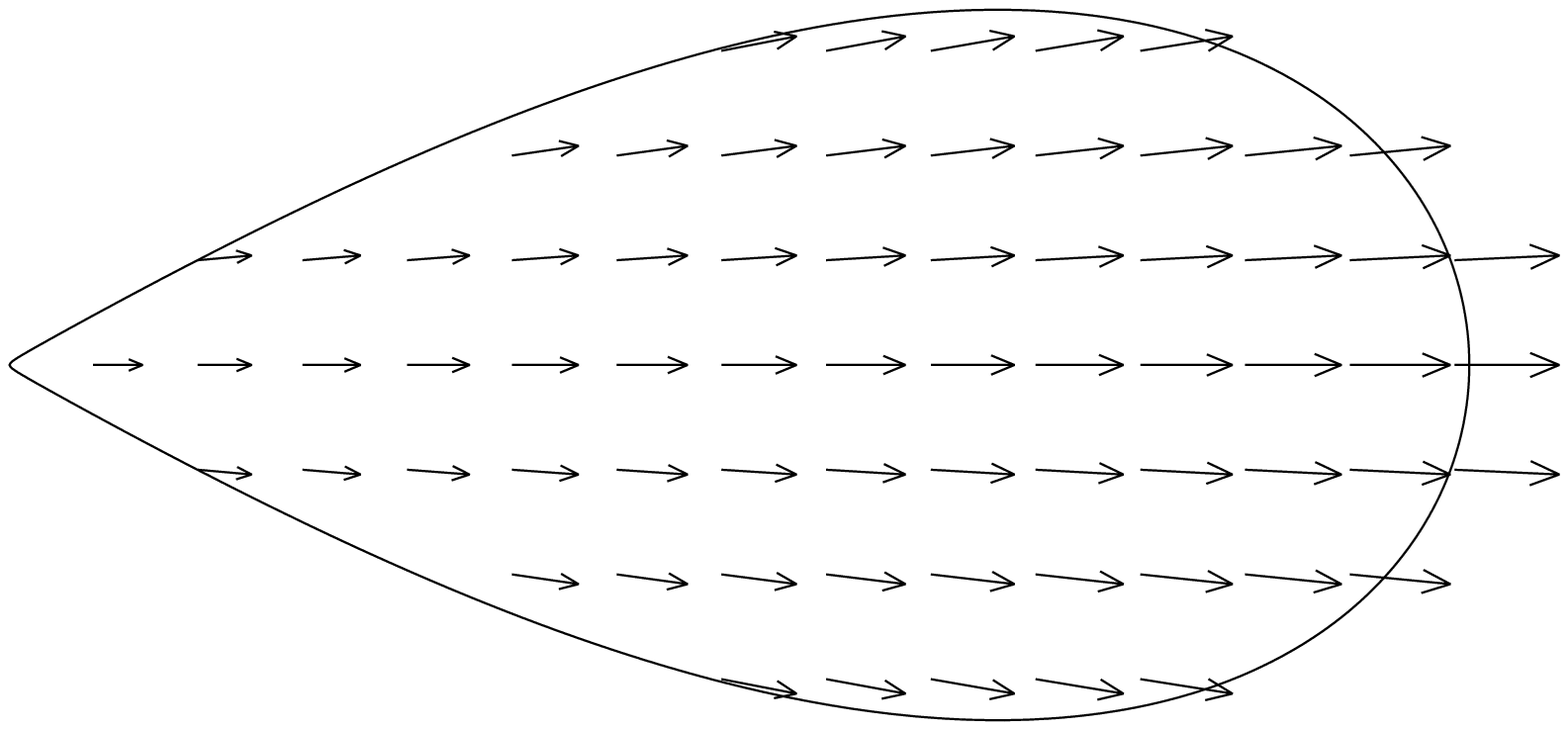}{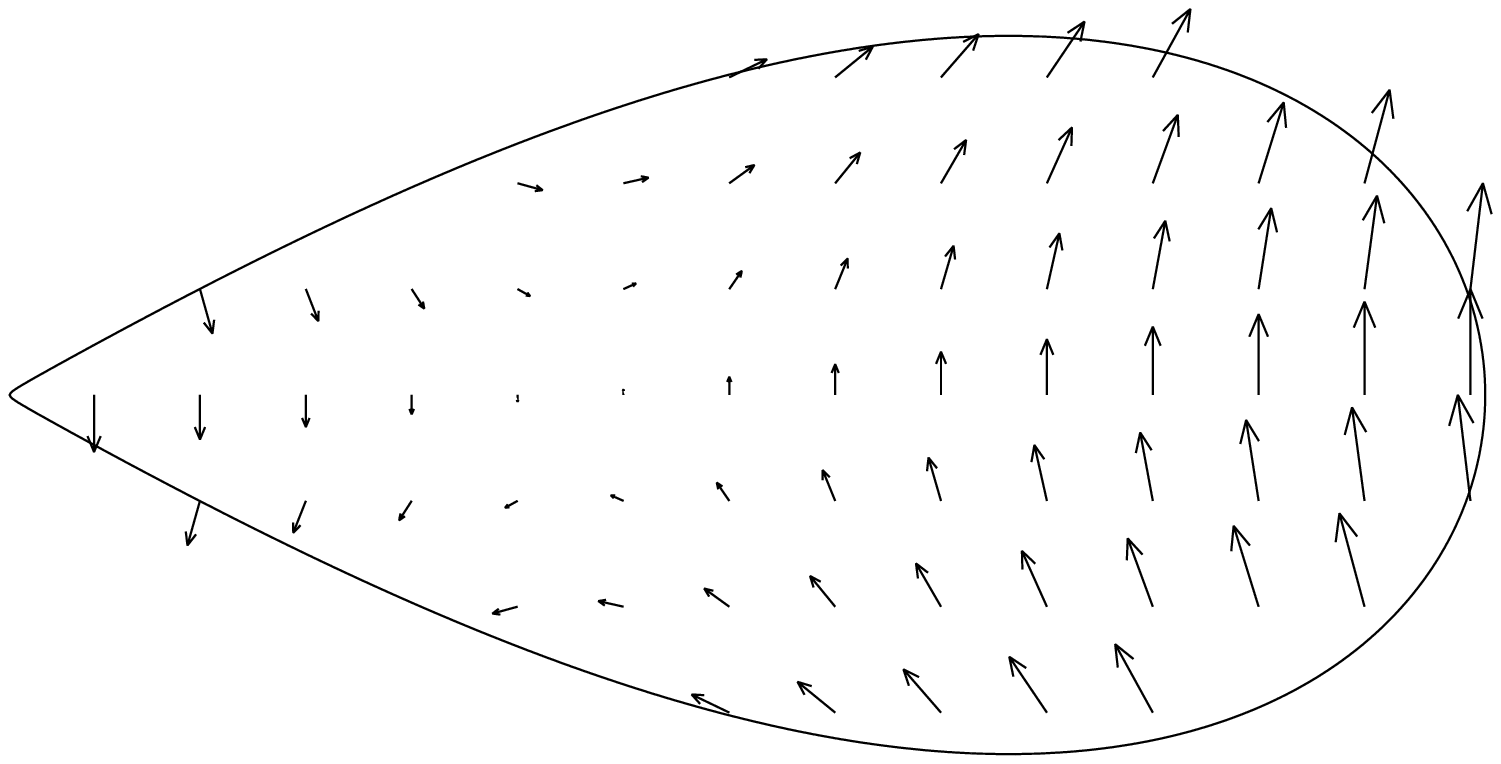}
\caption{Poloidal velocity field for the radial
epicyclic mode (left) and the vertical epicyclic mode (right)
of a nonslender $n=3$ torus with $\beta=0.134589$ and
pressure maximum at $r=7.293M$ (the same torus illustrated in the right hand
panel of Fig.~\ref{3.2}).}
\label{modevelocities}
\end{figure}

The analytic, poloidal velocity fields of the axisymmetric radial and
vertical epicyclic modes of this nonslender torus are shown in Figure
\ref{modevelocities}.  The radial mode involves some vertical expansion
on outward displacements (and corresponding vertical compression on inward
displacements).  The reason for this is that the vertical tidal gravity
is less at larger radii, and so the torus expands under vertical
pressure gradients.  For the vertical epicyclic mode, the motions of at
least the outer parts of the torus are also easy to understand.  The
upper half of the torus moves radially outward (and the lower half moves
radially inward) on upward vertical displacements.  This is because the
radial component of the gravitational field decreases as one moves off
the midplane, and so centrifugal and pressure gradient accelerations
drive outward radial displacements in the upper half of the torus.
Similarly the increasing gravity as the lower half moves toward the
midplane causes inward radial displacements.  The pressure forces
exerted by these radial motions act to oppose the vertical motions of
the inner parts of the torus.  This effect is so strong that
the inner parts of the torus shown in the Figure are actually oscillating
in the opposite direction to the zeroth order vertical epicyclic mode.
Our perturbation expansion may therefore not be accurate for this thick
a torus.  It nevertheless makes sense that the
vertical displacements in the vertical epicyclic mode of a radially
extended torus cannot in general all be in the same direction
simultaneously, as there are no forces that would maintain such radial
coherence.

Increasing the thickness of the torus (increasing $\beta$) always
decreases the axisymmetric mode frequencies so that they are less
than the test particle frequencies.  Our complicated analytic formulas
for the mode frequencies suggest that the physical reasons for the
details of this decrease must itself be complicated.  However, it is
not hard to guess physically why the mode frequencies should decrease
in general.  First, unless the radius of the pressure maximum is very
close to the innermost stable circular orbit, the center of mass of a
nonslender torus always lies outside the radius of the pressure maximum.
At large radii both of the test particle frequencies always decrease with
radius.  Hence the outward shift of the center of mass of the torus as
it thickens should alone decrease the frequency of the epicyclic modes.
At small radii, the test particle radial epicyclic frequency increases
with radius, and so one might expect that the radial epicyclic mode
frequency would increase with torus thickness (except perhaps when the
torus is very close to the innermost stable circular orbit and the center
of mass then lies inside the pressure maximum).  However, in this regime
the mode frequency appears to depend primarily on the nearby presence of
the cusp in the equipotential surfaces.  There is no radial restoring force
at the cusp, and as a result the radial epicyclic mode frequency of a
nonslender torus near the cusp drops dramatically, as shown in Figure
\ref{axisym}.  Finally, even when the pressure maximum of the torus is
close to the innermost stable circular orbit $r\lesssim6.5M$ so that the
center of mass of the torus moves inward on thickening, the vertical
epiyclic mode is still biased toward the outer parts of the torus because
pressure forces reduce the vertical oscillations in the inner parts.
As a result, the frequency of the vertical epicyclic mode still decreases
with torus thickness even in this regime.

%
              \subsubsection{Nonaxisymmetric Modes with $m=\pm1$}
%

Recently, \citet{bur06} suggested a new pair of modes having
the 3:2 commensurability, that would better satisfy the observational
constraints on black hole spins and masses, in particular 
the most recent spin estimates from spectral fitting to the X-ray 
spectrum of GRO
J1655-40 \citep{sha06,dav06}. These modes are the axisymmetric
vertical and the nonaxisymmetric $m=-1$ radial mode, whose test
particle frequencies occur in a 3:2 ratio close to the marginally
stable orbit. 

\begin{figure}
\plotone{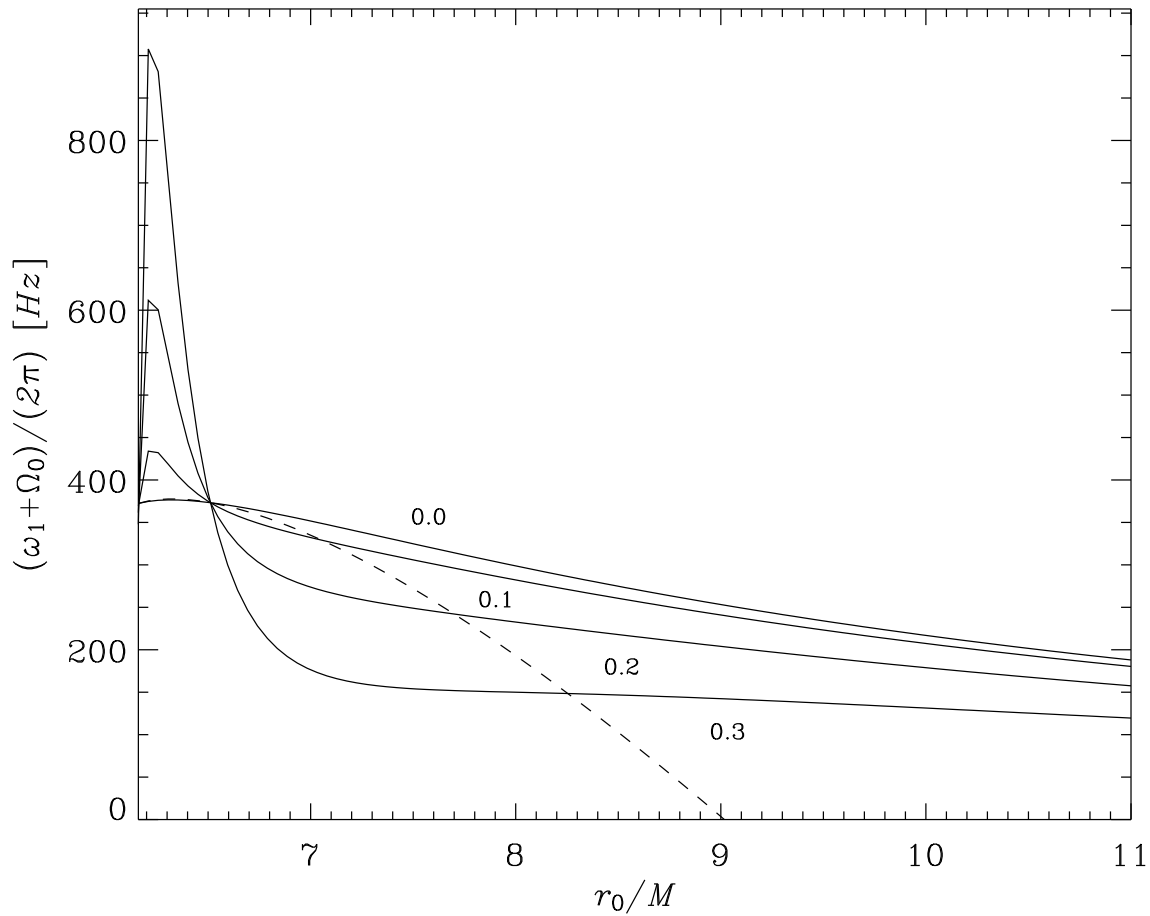}
\caption{The sum of the axisymmetric radial epicyclic mode frequency
$\omega_1$ and the orbital frequency $\Omega_0$ at the pressure maximum,
for an $n=3$
pseudo-Newtonian torus around a $M=10M_{\odot}$ black hole, as a function
of $r_0$
for different values of $\beta$. Physical tori exist only to the right
of and above the dashed line.}
\label{nonaxi1}
\end{figure}

\begin{figure}
\plotone{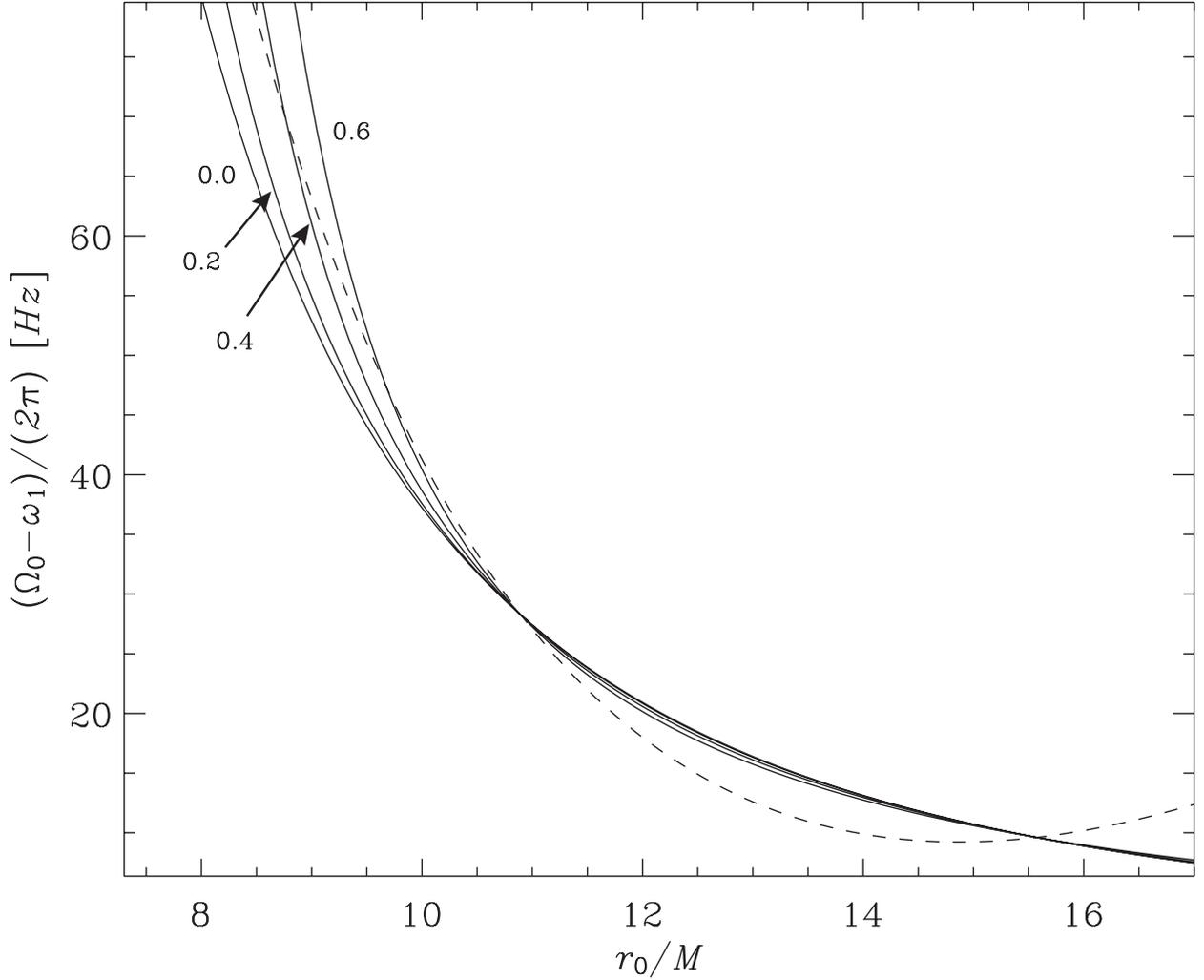}
\caption{The difference between the orbital frequency $\Omega_0$ at the
pressure maximum $r_0$ and the axisymmetric radial
epicyclic mode frequency for an $n=3$ pseudo-Newtonian torus around a
$M=10M_{\odot}$ black hole, as a function of $r_0$ for different values
of $\beta$. Between
the radii $r=6M$ and $r=10.895M$ the frequency difference increases
with increasing $\beta$, and physical tori exist only below the dashed
line. From $r=10.985M$ to $r=15.573 M$, the frequency difference
decreases slightly with increasing $\beta$, and physical tori
exist only above the dashed line.
For radii greater than $15.573M$ the behavior is similar to that of the
first interval.}
\label{nonaxi2}
\end{figure}

The nonaxisymmetric ($m=\pm1$) radial mode frequencies are shown in
Figures \ref{nonaxi1} and \ref{nonaxi2}. Again, the curves with
$\beta=0$ correspond to the frequency of a test particle, and the
dashed line denotes the limit for tori in equilibrium. Figure
\ref{nonaxi1} shows the sum of the axisymmetric radial frequency and
$\Omega_0$ (i.e., the $m=1$ mode) for different torus thicknesses. 
With increasing values of $\beta$, the sum of the frequencies slightly
decreases. The situation becomes more complicated 
for the $m=-1$ mode.
The frequency difference between $\Omega_0$ and the axisymmetric radial
epicyclic frequency first increases with growing $\beta$, but
then it decreases with
increasing $\beta$ in the interval $10.895M<r<15.573M$ (Figure \ref{nonaxi2}). 
Finally, for radii greater than $15.573M$, the frequency difference increases
with $\beta$ again.

\begin{figure}
\plottwo{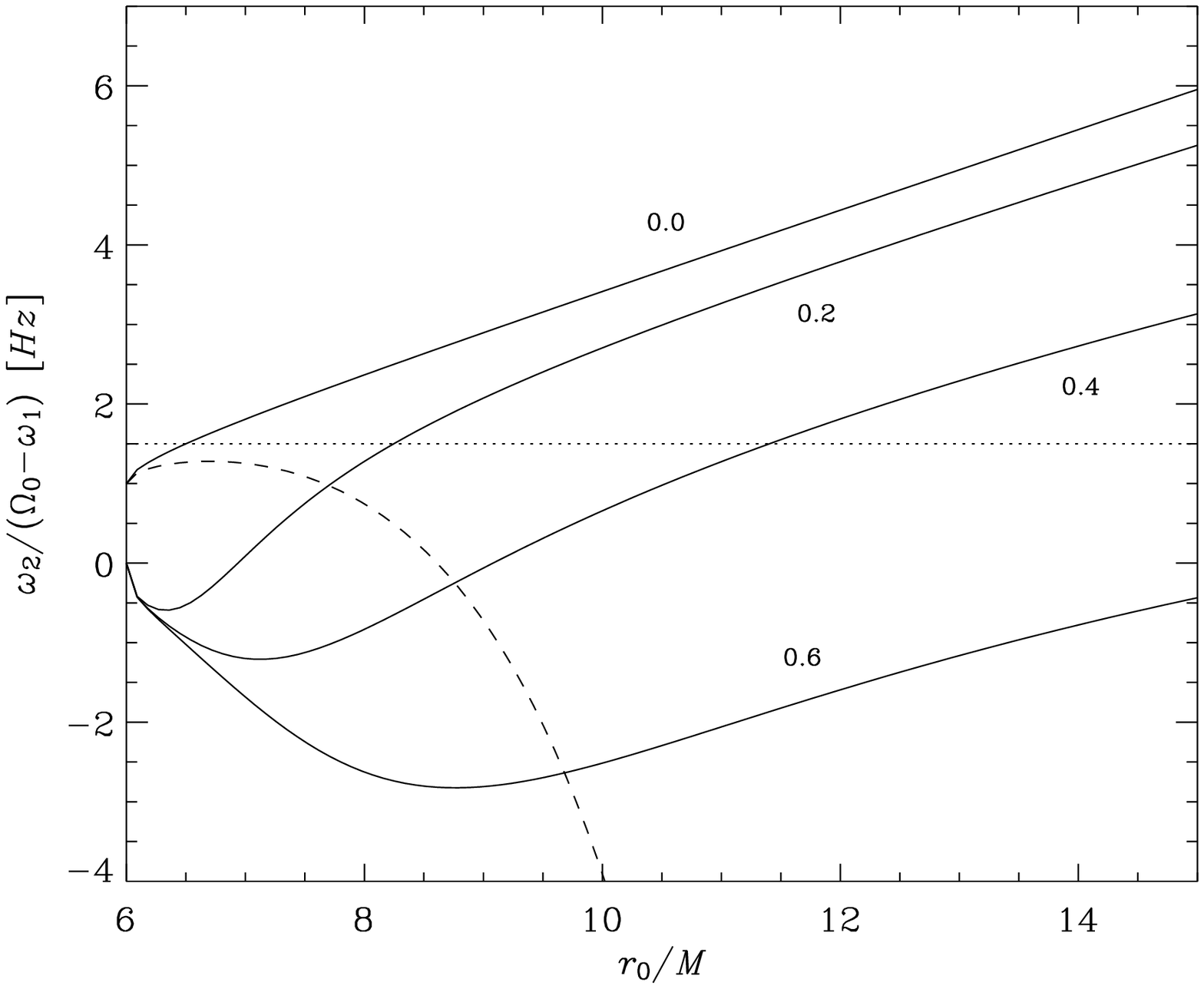}{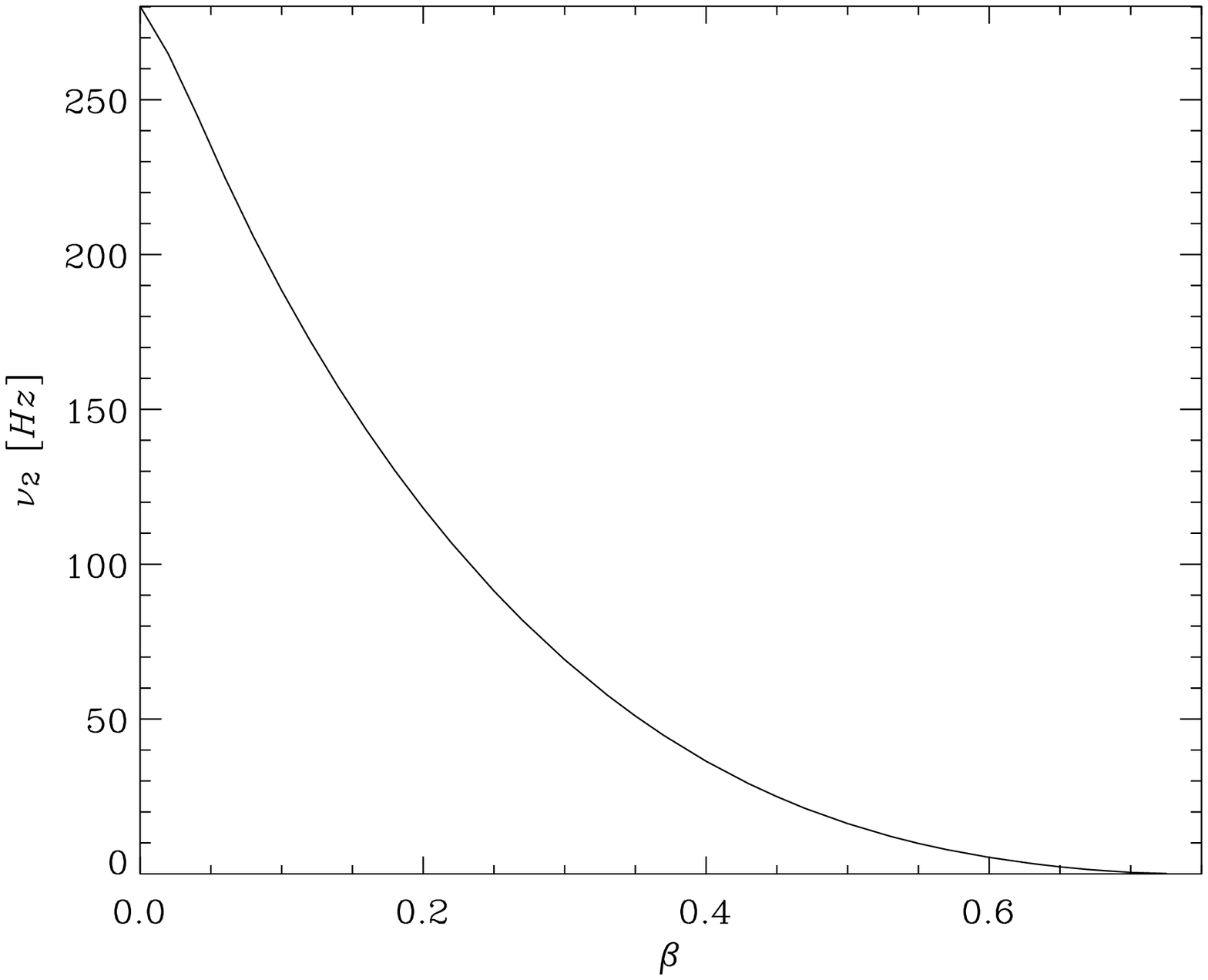}
\caption{{\it Left}: The ratio $\omega_2/(\Omega_0-\omega_1)$ of the
axisymmetric vertical mode frequency to the frequency of the $(m = -1)$
radial mode
for an $n=3$ torus in a pseudo-Newtonian potential, plotted for
different values of $\beta$ as a function of radius of the torus
pressure maximum. Physical models lie above the dashed line.
For all thick tori having $\beta < \beta_{max}=0.725108$,
there is one possible location of the torus pressure maximum at which
the modes are in a 3:2 ratio. For tori with $\beta \ge \beta_{max}$,
no such pressure maximum radii can exist. {\it Right}: The axisymmetric
vertical epicyclic mode frequency $\nu_2=\omega_2/(2\pi)$ (for the
same tori as in the left panel) at the
radius where $\omega_2/(\Omega_0-\omega_1)=3/2 $, plotted as a function
of $\beta$. At $\beta = \beta_{max}=0.725108$, the
frequency  $\nu_2$ goes to zero.}
\label{nonaxi3.2} 
\end{figure}

The ratio of the axisymmetric vertical frequency to the $m = -1$ 
radial mode frequency for different
torus thicknesses can be seen in the left panel of Figure \ref{nonaxi3.2}.
Unlike the resonance for the axisymmetric epicyclic 
modes, the radius at which these two modes are in a 3:2 ratio moves outward
with increasing $\beta$. This corresponds
to a decrease of the individual frequencies, as can be seen in the right
part of the Figure, 
and the frequencies eventually become zero for $\beta = 
\beta_{max}=0.725108$. For tori
with $\beta < \beta_{max}$, there exists exactly one torus pressure
maximum location that generates a 3:2 frequency ratio, and no
such location exists for still thicker tori corresponding to $\beta
\ge \beta_{max}$. This would suggest that even fairly large
tori can exhibit a 3:2 resonance, but the perturbative
method is valid only for small $\beta$, and any firm results on very 
extended tori must be obtained by other methods.

%
              \subsection{Consequences for Black Hole Spin Estimates}
%

The first black hole spin estimates from fitting the
axisymmetric epicyclic mode frequencies to the observed QPOs from 
GRO J1655-40 were reported by \citet{abr01}. Recently \citet{tor05} 
have used the orbital resonance model to estimate the spins of all 
three microquasars with known masses.  The most tightly constrained
spin was in GRO~J1655-40, where $a/M$ was found to lie between 0.93
and 0.99.  This spin value that was derived from a resonance between the
axisymmetric epicyclic modes is not in accord with the independent
spin measurements, determined by fitting the X-ray spectral
continua: $a/M\sim0.65-0.75$ \citep{sha06}, $a/M\sim0.62$ or less if
the disk inclination is free \citep{dav06} for
GRO~J1655-40.

However, the resonance model estimates were based on 
the epicyclic frequencies for free test particles. Figure \ref{3.2} 
shows that the axisymmetric epicyclic frequencies at the resonant 
radius will be higher for a nonslender torus than for a test 
particle. The maximal frequency shift due to pressure effects 
is around 15 percent of the test particle frequency. 
The previous studies have therefore most probably 
{\it overestimated} the actual value of the spin of the 
black hole, which explains some of the discrepancies with
other methods.

On the other hand a resonance between the axisymmetric vertical
epicyclic mode and the nonaxisymmetric $m=-1$ radial mode
yields a black hole spin for GRO J1655-40 \citep{bur06}
compatible with other recent spin estimates although Bursa 
used the frequencies for free particles. As seen from Figure 
\ref{nonaxi3.2}, the frequencies of these
two modes at the resonant radius decrease with growing torus
thickness, contrary to the axisymmetric modes.  Therefore in this case
the black hole spins should  be $\it higher$
than previously estimated. The maximal shift of the frequencies
(and consequently the black hole spin) can be
very large,  but our perturbative method is valid for small $\beta$,
and is not reliable for $\beta \approx 0.7$.

The above modifications of the spin estimates are
based on purely Newtonian calculations using the
Paczy{\'n}ski-Wiita gravitational potential. In order to explore the
real behavior of thick tori orbiting around
rotating black holes, one needs to solve the hydrodynamical equations
in the Kerr metric.

%
              \subsection{Comparison with Numerical Simulation}
%

We compare our analytically calculated frequencies for the axisymmetric
epicyclic modes with those from hydrodynamic numerical simulations.
The numerical simulations were performed using the
ZEUS-2D code by \citet{sto92}. The
radial, and vertical epicyclic oscillations, respectively,
were excited by applying
a purely radial, or vertical constant velocity
field perturbation, respectively, to the torus at t=0. Then we 
Fourier analyzed the
radial and vertical positions of the center of mass
in order to find the oscillatory frequencies. 

\begin{figure}
\plottwo{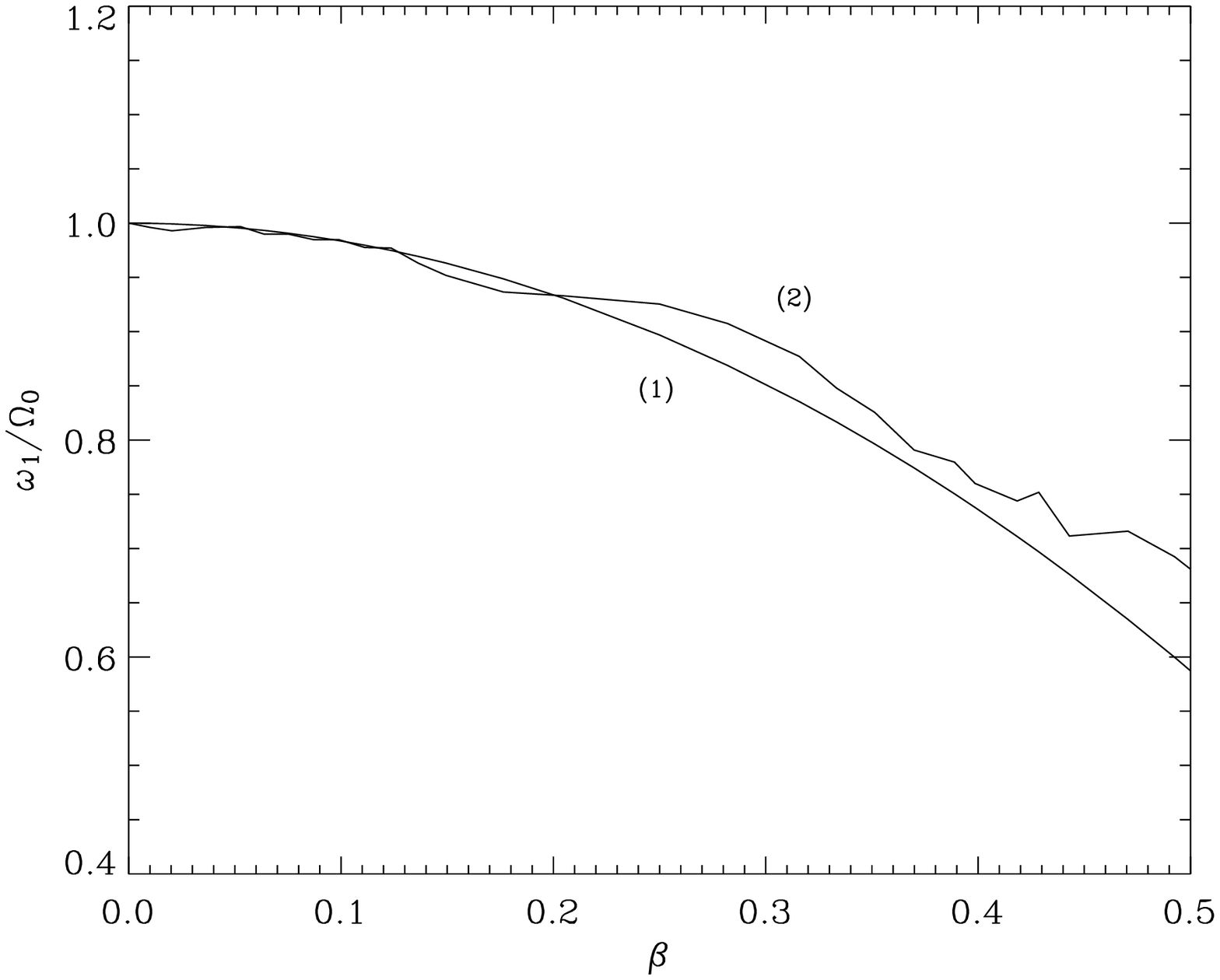}{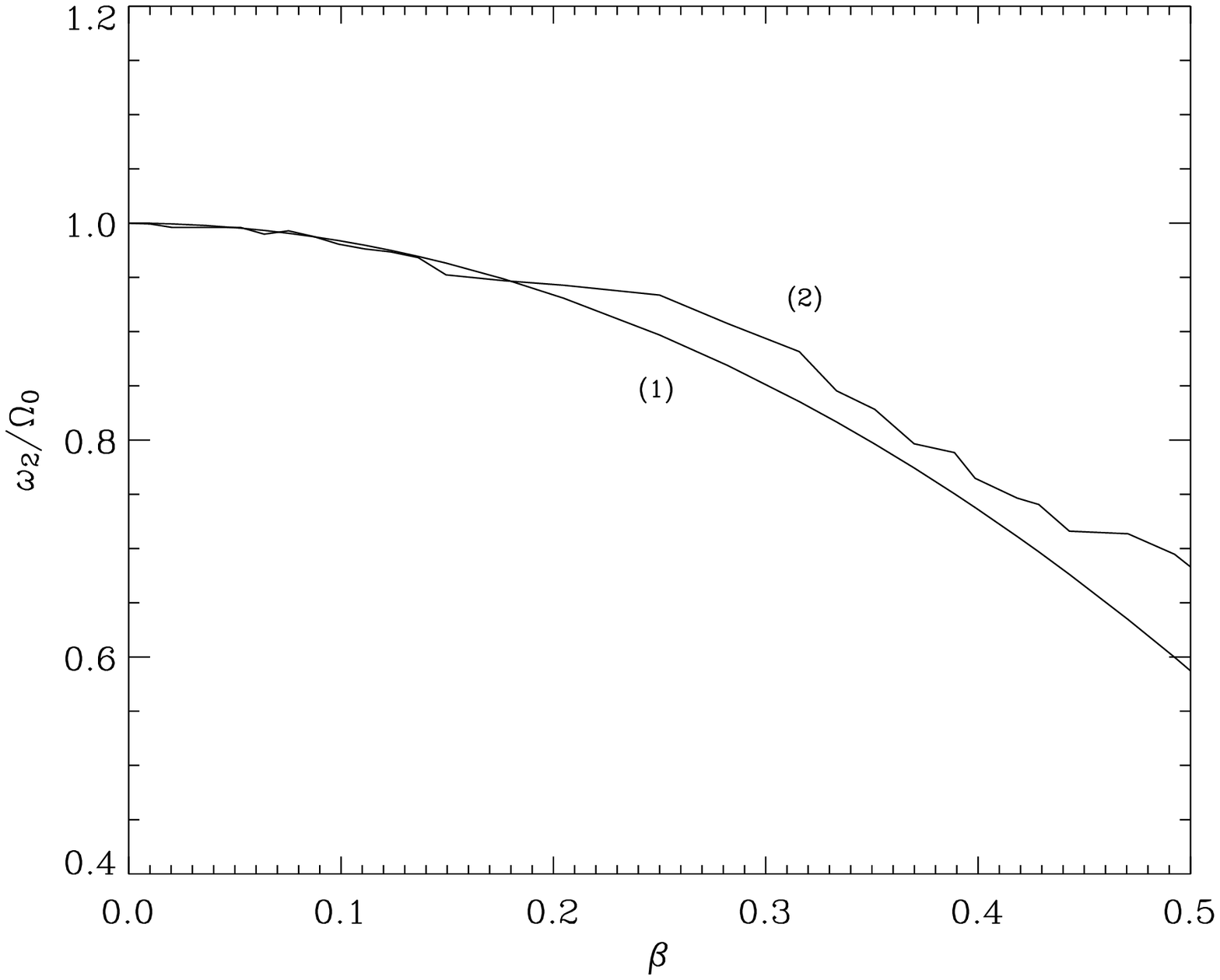}
\caption{{\it Left}: Comparison between the analytically (1) and
numerically (2) calculated axisymmetric radial epicyclic mode
frequency for an $n=3$ torus orbiting in a spherically symmetric point
mass potential, plotted as a function of $\beta$. {\it Right}: Same as
the left panel, but for the axisymmetric vertical epicyclic mode.}  
\label{newton} 
\end{figure}

\begin{figure}
\plottwo{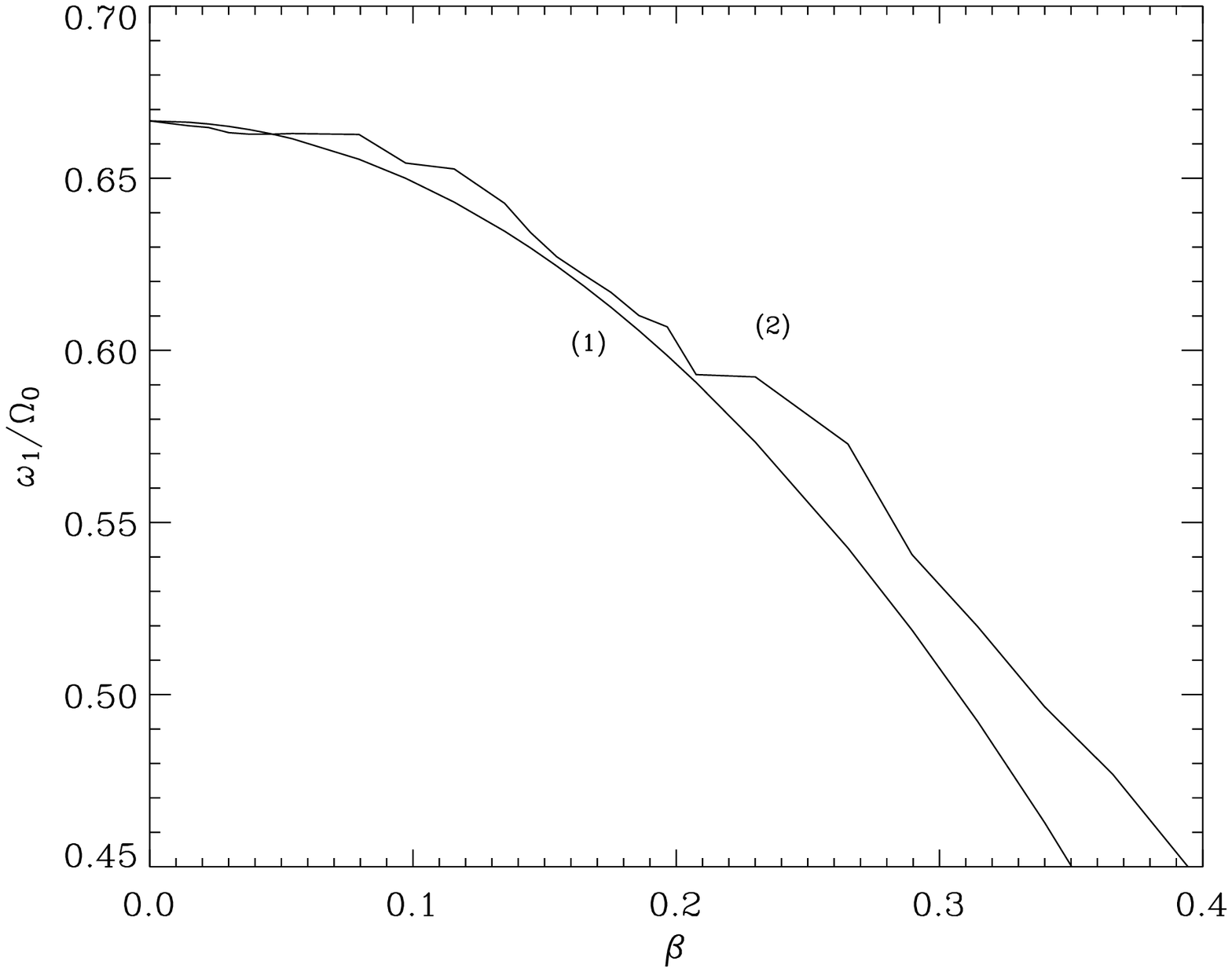}{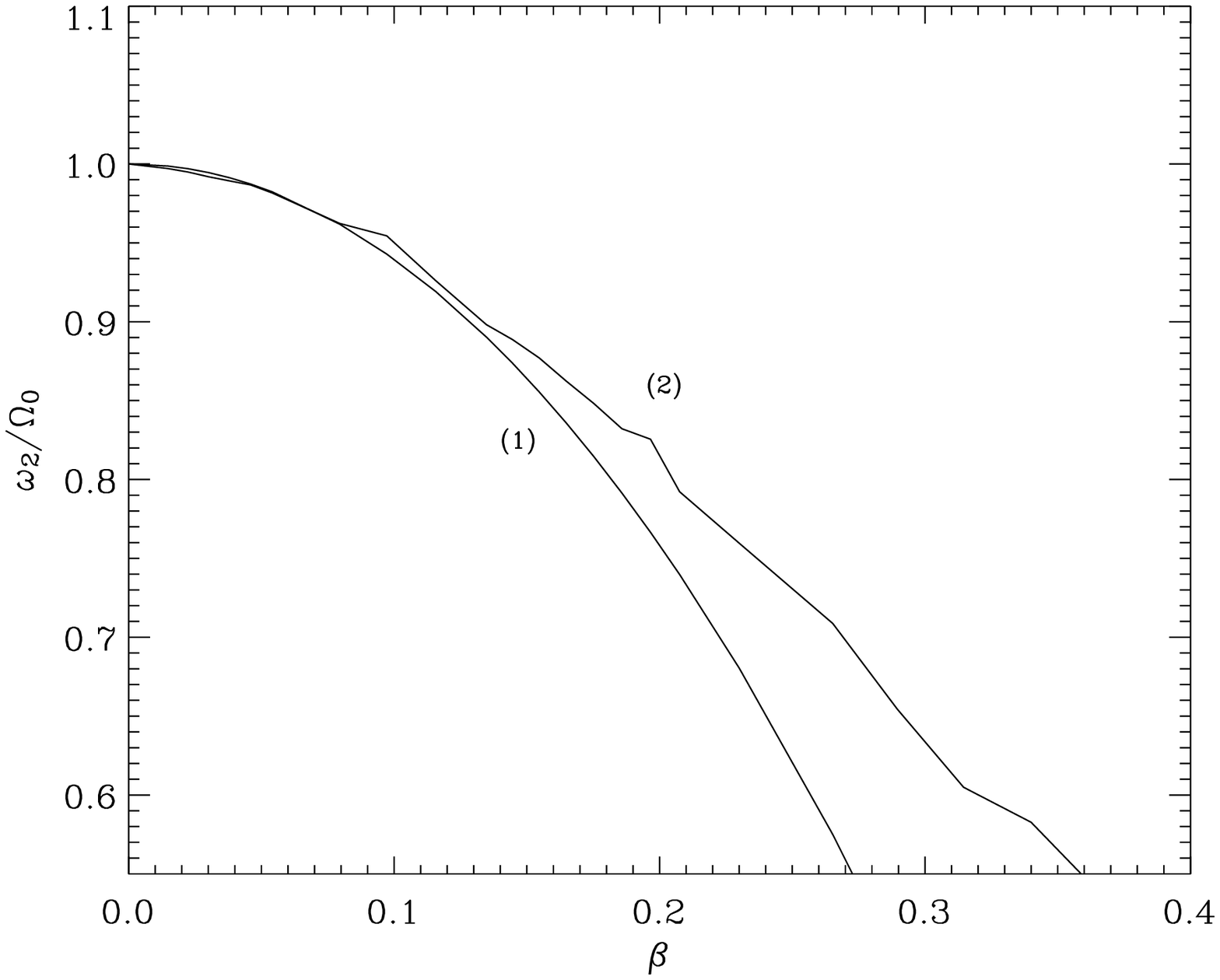}
\caption{{\it Left}: Comparison between the analytically (1) and
numerically (2) calculated axisymmetric radial epicyclic mode
frequency for an $n=3$ torus orbiting in a pseudo-Newtonian potential,
plotted as a function of $\beta$. The pressure maximum of the torus is
at $9.2 M$. For this radius, the maximum value of $\beta$ for which an
equilibrium torus can exist, is 0.492968. {\it Right}: Same as
the left panel, but for the axisymmetric vertical epicylic mode.}
\label{pseudo}
\end{figure}

Figure \ref{newton} compares the results of our analytic
expressions (\ref{omradnewt}) and (\ref{omvertnewt}) with $m=0$ to the
epicyclic mode frequencies measured from the simulations 
as a function of torus
thickness. The agreement is excellent for $\beta\lta0.2$.  Figure
\ref{pseudo} similarly compares the analytic $m=0$ frequencies from
our perturbation theory to the simulation results for tori in
pseudo-Newtonian potentials.  Once again, there is good agreement for
$\beta\lta0.2$.
The general trend that the frequencies decrease with increasing $\beta$
was also observed by \citet{rub05} in their simulations.

%
              \section{Discussion and Conclusions}
	      \label{sec:conclusions}
%

In this paper we have assumed that (possibly nonaxisymmetric) vertical
and radial epicyclic modes of fluid tori are responsible for the
commensurate pairs of high frequency QPOs in black hole X-ray binaries.
We have derived exact analytic formulae for the mode eigenfunctions
and frequencies.  We find that the frequencies are always below those
corresponding to free test particles, an effect that must be taken
into account before identifying such modes with observed QPOs.

While our analytic calculations are exact, they apply to very simplified
and idealized configurations:

(1) Our calculations were restricted to Newtonian mechanics.  This is a
necessary first step to doing the fully general relativistic calculation
in Kerr spacetime (Straub et al. 2007, in preparation).  Moreover,
Newtonian potentials that mock up various aspects of the physics of
Schwarzschild spacetime are used by many numericists (e.g.
\citealt{lee04,mac06}), and our results may be applied directly
to their simulations.

(2) We considered isolated, non-accreting, non-magnetized, polytropic tori.
The flow configuration of the accretion disks in X-ray binaries when they
exhibit QPOs is still far from clear, but pressure-supported torus-like
configurations remain an interesting possibility.  Such configurations are
seen as ``inner tori'' in global MRI simulations \citep{dev03,mac06,mat07}.
These inner tori have magnetic energy densities much less than the gas
internal energy density, so magnetic fields are less important to the overall
hydrostatic structure than gas pressure.  (Strongly magnetized configurations
may however exist in real systems, and our treatment in this paper would not
be valid for them.)  Among all the possible modes of an isolated torus, the
epicyclic modes will be most robust to boundary conditions, as they correspond
to physical displacements of the entire torus.  We therefore believe that our
calculations for the corrections to the mode frequencies will be reasonably
robust to these boundary conditions.  The entropy distribution of these tori
is completely unknown, and existing simulations tell us little as they do
not have consistent thermodynamics.  Nevertheless, the luminous accretion
flows exhibiting high frequency QPOs are likely to be radiation pressure
supported, and a polytropic configuration with $n=3$ may therefore
be fairly close to reality.

(3) For mathematical
reasons, we also restricted ourselves to constant specific angular momentum
tori.  Such angular momentum distributions appear unlikely to exist in
nature, as global simulations of disks with MRI turbulence produce inner
tori with angular momentum distributions which are closer to the test
particle (``Keplerian'') limit.  (Note, however, that these simulations
currently lack optically thick radiative cooling which is likely to
determine the overall pressure support of the flow, and therefore departures
from Keplerian angular momentum.)  The Papaloizou-Pringle equation with
non-constant specific angular momentum is not a self-adjoint eigenvalue
problem, and so we do not have a complete set of orthogonal basis eigenfunctions
in order to expand the perturbed modes around the slender torus limit.
The perturbation equations using the Lagrangian displacement
vector as dependent variable can be used to generate a complete set of
orthogonal eigenfunctions, at least within Newtonian mechanics \citep{sch01},
and this may enable an analytic calculation of the corrections to the mode
frequencies for non-constant specific angular momentum tori.  We note that
fully general relativistic simulations in Kerr spacetime of non-constant
specific angular momentum oscillating tori exhibit axisymmetric radial
and vertical epicyclic mode frequencies which are below the test particle
frequencies (Fragile 2006, private communication), in agreement with the
qualitative results we have found here.

We view the results of the calculations presented here as a necessary
tool for identifying the modes in fully 3D, time dependent numerical
simulations of magnetohydrodynamic tori. A preliminary example of what
can be achieved is provided by \citet{bur06}, who used our results to
identify the $(m = 1)$ radial and vertical epicyclic modes in the 
recent numerical simulations by \citet{mac06}.  Once identified, our
analytic expressions for the mode eigenfunctions may also be useful
in trying to explain why such modes are excited in simulations.  They
may also be useful in making predictions for the observed X-ray
modulation.  As explored numerically by \citet{bur04} and \citet{sch06},
gravitational redshifts, Doppler shifts, and lensing of X-ray photons
emitted by tori oscillating in both the vertical and radial epicyclic
modes can produce detectable oscillations in the predicted X-ray fluxes
measured by an observer.

\begin{figure}
\plotone{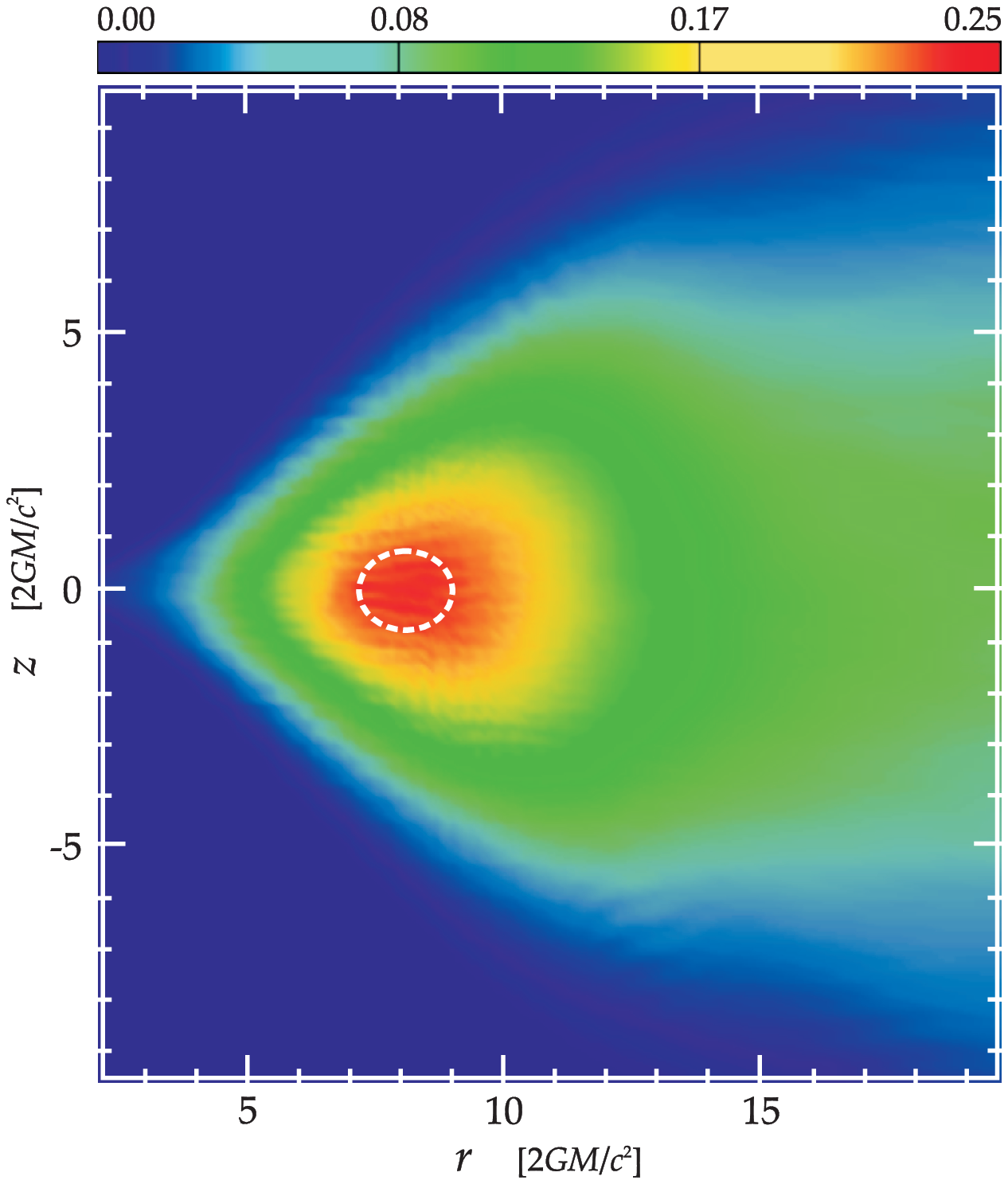}
  \caption{A pressure supported inner torus is apparent in the
  MHD simulations described by \citet{mac06} and \citet{mat07}.
  The relative density is 
  color coded. The superimposed elliptical shape (a white broken line)
  corresponds to the analytic nonslender torus studied in this paper.}
  \label{fig-mami-torus}
\end{figure}

%
              \acknowledgments
{We thank Mami Machida and Michal Bursa for preparing 
Figure~\ref{fig-mami-torus} for us. We very much thank these colleagues
and also Didier Barret, Axel Brandenburg, Chris Fragile, 
Ji{\v r}\'i Hor{\'a}k, Vladim\'ir Karas, Shoji Kato, Michiel van der Klis, 
Jean-Pierre Lasota, William Lee, Ryoji Matsumoto, Jeff McClintock, 
Mariano Mend{\'e}z, Paola Rebusco, Ron Remillard, Giora Shaviv, 
Zden{\v e}k Stuchl{\'{\i}}k, Gabriel T\"or\"ok, Roberto Vio, Bob Wagoner and 
Piotr {\.Z}ycki for comments, criticism, and several helpful 
suggestions. Our work was supported in part by the following grants: 
NSF's PHY99-07949 and AST03-07657, Nordita's 2005 Nordic Project awarded 
to M.A.A., Czech MSM~4781305903 and LC06014, and Polish 1P03D~005~30. 
The authors are also grateful to the following institutions for hosting them 
while much of this work was carried out: the Kavli Institute for 
Theoretical Physics in Santa Barbara, the Institut d'Astrophysique 
de Paris, Nordita in Copenhagen, and the N.~Copernicus Astronomical 
Centre in Warsaw.}

%

%
%

%
%

%
%

%
%

%
%

%
%

%
%

%
%

%
%

%
%

\vfill\eject

%
%
\begin{deluxetable}{ccc}
\rotate
\tabletypesize{\scriptsize}
\tablecaption{Simplest Modes of the General, Constant Specific Angular Momentum
Slender Torus. \label{tbl-1}}
\tablewidth{0pt}
\tablehead{
\colhead{$J$\tablenotemark{a}} &\colhead{$\bar{\sigma}_0^2$} &
\colhead{Eigenfunction\tablenotemark{b}}
}
\startdata
0 & 0 & $a_0$ \\
1 & $\bar{\omega}_r^2$ & $a_1\bar{x}$ \\
2 & $\baromz^2$ & $a_2\bar{y}$ \\
3 & $\bar{\omega}_r^2+\baromz^2$ & $a_3\bar{x}\bar{y}$ \\
4 & ${(2n+1)(\bar{\omega}_r^2+\baromz^2)-[4n(n+1)
     (\baromz^2-\bar{\omega}_r^2)^2+(\bar{\omega}_r^2+
     \baromz^2)^2]^{1/2}\over2n}$ & 
    $a_4\left\{1+{n\bar{\sigma}_0^2\over\baromz^2-\baromr^2}(\baromr^2
     \bar{x}^2-\baromz^2\bar{y}^2)-{2(n+1)\baromz^2\baromr^2\over\baromz^2-
     \baromr^2}(\bar{x}^2-\bar{y}^2)\right\}$ \\
5 & ${(2n+1)(\bar{\omega}_r^2+\baromz^2)+[4n(n+1)
    (\baromz^2-\bar{\omega}_r^2)^2+(\bar{\omega}_r^2+
    \baromz^2)^2]^{1/2}\over2n}$ &
    $a_5\left\{1+{n\bar{\sigma}_0^2\over\baromz^2-\baromr^2}(\baromr^2
    \bar{x}^2-\baromz^2\bar{y}^2)-{2(n+1)\baromz^2\baromr^2\over\baromz^2-
    \baromr^2}(\bar{x}^2-\bar{y}^2)\right\}$ \\
\enddata

\tablenotetext{a}{$J$ is an arbitrary nonnegative integer index which we are
using to label the modes.  A more physically descriptive set of labels can be
found in the paper by \citet{bla06}.}
\tablenotetext{b}{The constants $\{a_0, a_1, ..., a_{5}\}$ are given in
Table 2, and are chosen such that the eigenfunctions are normalized in the
inner product (\ref{ipslender}).}
\end{deluxetable}

%
%
\begin{deluxetable}{cc}
\rotate
\tabletypesize{\scriptsize}
\tablecaption{Normalization Constants for the Eigenmodes of Table 1.
\label{tbl-2}}
\tablewidth{0pt}
\tablehead{
\colhead{$J$} &\colhead{$a_J$}
}
\startdata
0 & $\left({n\baromr\baromz\over\pi}\right)^{1/2}$ \\
1 & $a_0[2(n+1)\baromr^2]^{1/2}$ \\
2 & $a_0[2(n+1)\baromz^2]^{1/2}$ \\
3 & $a_0[4(n+1)(n+2)\baromr^2\baromz^2]^{1/2}$ \\
4 & $a_0\left\{{(n+2)[\bar{\sigma}_0^2-(\baromz^2+
     \bar{\omega}_r^2)]\over2n\bar{\sigma}_0^2-(2n+1)(\baromz^2
     +\bar{\omega}_r^2)}\right\}^{1/2}$ \\
5 & $a_0\left\{{(n+2)[\bar{\sigma}_0^2-(\baromz^2+
     \bar{\omega}_r^2)]\over2n\bar{\sigma}_0^2-(2n+1)(\baromz^2
     +\bar{\omega}_r^2)}\right\}^{1/2}$ \\
\enddata
\end{deluxetable}


\begin{thebibliography}{}
%
\bibitem[Abramowicz(2005b)]{abr05}Abramowicz, M.A., 2005, 
AN, 326, 782
\bibitem[Abramowicz et al.(2006)]{abr06}Abramowicz, M.A., 
Blaes, O.M., Hor\'ak, J., Klu\'zniak, W. \& Rebusco, P., 2006, 
Classical and Quantum Gravity, 23, 1689 
\bibitem[Abramowicz \& Klu{\'z}niak (2001)]{abr01}Abramowicz, M.A. 
\& Klu\'zniak, W., 2001,
A\&A, 374, L19
\bibitem[Blaes(1985)]{bla85}Blaes~O.~M., 1985, 
MNRAS, 216, 553
\bibitem[Blaes, Arras \& Fragile(2006)]{bla06}Blaes, O.M., Arras, P.
\& Fragile, P.C., 2006, 
MNRAS, 369, 1235
\bibitem[Bursa (2006)]{bur06}Bursa, M., 2006, 
Ph. D. thesis, Charles Univ., Prague
\bibitem[Bursa et al.(2004)]{bur04}Bursa, M., Abramowicz, M.A., 
Karas, V. \& Klu\'zniak, W., 2004, 
ApJ, 617, L45
\bibitem[Davis, Done, \& Blaes(2006)]{dav06}Davis, S. W., Done, C.,
\& Blaes, O. M. 2006, ApJ, 647, 525
\bibitem[De Villiers, Hawley \& Krolik(2003)]{dev03}De Villiers, J.-P.,
Hawley, J. F., \& Krolik, J. H., 2003, ApJ, 599, 1238
\bibitem[Hawley \& Balbus(2002)]{haw02}Hawley, J. F., \& Balbus,
S. A. 2002, ApJ, 573, 738
\bibitem[Klu{\'z}niak \& Abramowicz(2001a)]{klu01a}Klu\'zniak, W. 
\& Abramowicz, M.A., 2001a, 
Phys. Rev. Lett., submitted [astro-ph/0105057]
\bibitem[Klu{\'z}niak \& Abramowicz(2001b)]{klu01}Klu\'zniak, W., \&
Abramowicz, M. A. 2001b, Acta Phys. Pol. B, B32, 3605
\bibitem[Klu{\'z}niak \& Abramowicz(2002)]{klu02}Klu\'zniak, W.
\& Abramowicz, M.A., 2002, A\&A, submitted [astro-ph/0203314] 
\bibitem[Lee, Abramowicz \& Klu\'zniak(2004)]{lee04}Lee, W.H., 
Abramowicz, M.A. \& Klu\'zniak, W., 2004, 
ApJ, 603, L93
\bibitem[Machida et al.(2006)]{mac06}Machida, M., Nakamura K.E.
\& Matsumoto, R., 2006, PASJ, 58, 193
\bibitem[Matsumoto \& Machida(2007)]{mat07}Matsumoto, R. 
\& Machida, M., 2007, in IAU Symp. 238, Black Holes from Stars to
Galaxies, ed. V. Karas \& G. Mat (Cambridge: Cambridge Univ. Press),
37
\bibitem[Nayfeh \& Mook(1979)]{nay79}Nayfeh, A.H. 
\& Mook, D.A., 1979, 
Nonlinear Oscillations (New York: Wiley)
\bibitem[Paczy\'nski \& Wiita(1980)]{pac80} Paczy\'nski, B. 
\& Wiita, P.J., 1980, A\&A, 88, 23
\bibitem[Papaloizou \& Pringle(1984)]{pap84}Papaloizou, J.C.B. 
\& Pringle, J.E., 1984, 
MNRAS, 208, 721
\bibitem[Remillard \& McClintock(2006)]{rem06}Remillard, R. A., \&
McClintock, J. E. 2006, ARA\&A, 44, 49
\bibitem[Rezzolla et al.(2003)]{rez03} Rezzolla, L., Yoshida, S'i., 
Maccarone, T. J. \& Zanotti, O., 2003, 
MNRAS, 344, L37
\bibitem[Rubio-Herrera \& Lee(2005)]{rub05} Rubio-Herrera, E. 
\& Lee, W.H., 2005, 
MNRAS, 362, 789
\bibitem[Schenk et al.(2005)]{sch01}Schenk, A.K., Arras, P., 
Flanagan, E.E., Teukolsky, S.A. \& Wasserman, I., 2001, 
Phys. Rev. D, 65, 024001
\bibitem[Schnittman \& Rezzolla(2006)]{sch06}Schnittman, J. D., \&
Rezzolla, L. 2006, ApJ, 637, L113
\bibitem[Shafee et al.(2006)]{sha06}Shafee, R., McClintock, J. E., Narayan,
R., Davis, S. W., Li, L., \& Remillard, R. A. 2006, ApJ, 636, L113
\bibitem[Stone \& Norman(1992)]{sto92} Stone, J.M. 
\& Norman, M. L., 1992, 
ApJS, 80, 791
\bibitem[Strohmayer(2001)]{str01}Strohmayer, T. E., 2001, ApJ, 552,
L49
\bibitem[Tassoul(1978)]{tas78}Tassoul, J.-L., 1978, 
Theory of Rotating Stars, Princeton University Press, Princeton
\bibitem[T\"or\"ok et al.(2005)]{tor05} T\"or\"ok, G., 
Abramowicz, M.A., Klu\'zniak, W. \& Stuchl\'ik, Z., 2005, 
A\&A, 436, 1 
\bibitem[Zanotti, Rezzolla \& Font(2003)]{zan03} Zanotti, O., 
Rezzolla, L. \& Font, J.A., 2003, 
MNRAS, 341, 832

\end{thebibliography}
\end{document}